\documentclass[10pt,twocolumn,notitlepage,preprintnumbers,amsmath,amssymb,nofootinbib,superscriptaddress,longbibliography]{revtex4-1}

\usepackage{graphicx,longtable,mathrsfs,color,array}
\usepackage[hidelinks]{hyperref}
\usepackage[usenames,dvipsnames]{xcolor} 
\usepackage{amssymb,amsmath,mathtools,mathrsfs,slashed} 
\usepackage{epsfig,float} 
\usepackage{booktabs,longtable,ctable} 
\usepackage{exscale} 
\usepackage[normalem]{ulem} 
\usepackage{enumerate}
\usepackage{times, mathptmx} 
\usepackage{color}
\usepackage{hyperref}
\usepackage{graphicx}
\usepackage{color}
\usepackage{graphicx,graphics}
\usepackage{color}

\newcommand{\newText}[1]{{\color{black}{#1}}}

\newcommand{\hMpc}{\,h {\rm Mpc}^{-1}}
\newcommand{\Mpch}{\,{\rm Mpc}/h}
\newcommand{\lcdm}{\Lambda {\rm CDM}}

\newcommand{\refSec}[1]{Sec.~\ref{#1}}

\newcommand{\refFig}[1]{Fig.~\ref{#1}}
\newcommand{\refTable}[1]{Table~\ref{#1}}

\newcommand{\jcap}{JCAP}
\newcommand{\mnras}{MNRAS}
\newcommand{\apjs}{ApJs}
\newcommand{\physrep}{PhysRep}

\newcommand{\halofit}{\texttt{HALOFIT}\,}
\newcommand{\ecosmog}{\texttt{ECOSMOG}\,}
\newcommand{\mggadget}{\texttt{MG-GADGET}\,}
\newcommand{\mghalofit}{\texttt{MGHALOFIT}\,}

\newcommand{\camb}{\texttt{CAMB}\,}
\newcommand{\class}{\texttt{CLASS}\,}
\newcommand{\hmcode}{\texttt{HMCode}\,}
\newcommand{\powmes}{\texttt{POWMES}\,}
\newcommand{\euclidemu}{\texttt{EuclidEmulator}\,}
\newcommand{\cosmicemu}{\texttt{CosmicEmu}\,}

\begin{document}

\title{Emulators for the non-linear matter power spectrum beyond $\Lambda$CDM}

\author{Hans Winther}
\affiliation{Institute of Cosmology \& Gravitation, University of Portsmouth, Portsmouth, Hampshire, PO1 3FX, UK}
\affiliation{Institute of Theoretical Astrophysics, University of Oslo, 0315 Oslo, Norway}

\author{Santiago Casas}
\affiliation{AIM, CEA, CNRS, Universit\'e Paris-Saclay, Universit\'e Paris Diderot, Sorbonne Paris Cit\'e, F-91191 Gif-sur-Yvette, France}

\author{Marco Baldi}
\affiliation{Dipartimento di Fisica e Astronomia, Alma Mater Studiorum Università di Bologna, viale Berti Pichat, 6/2, I-40127 Bologna, Italy}
\affiliation{INAF - Osservatorio Astronomico di Bologna, via Ranzani 1, I-40127 Bologna, Italy}
\affiliation{INFN - Sezione di Bologna, viale Berti Pichat 6/2, I-40127 Bologna, Italy}

\author{Kazuya Koyama}
\affiliation{Institute of Cosmology \& Gravitation, University of Portsmouth, Portsmouth, Hampshire, PO1 3FX, UK}

\author{Baojiu Li}
\affiliation{Institute for Computational Cosmology, Department of Physics, Durham University, Durham DH1 3LE, U.K.}

\author{Lucas Lombriser}
\affiliation{Département de Physique Théorique, Université de Genève, 24 quai Ernest Ansermet, 1211 Genève 4, Switzerland}

\author{Gong-Bo Zhao}
\affiliation{National Astronomy Observatories, Chinese Academy of Science, Beijing, 100012, P.R.China}
\affiliation{Institute of Cosmology \& Gravitation, University of Portsmouth, Portsmouth, Hampshire, PO1 3FX, UK}

\date{Received \today; published -- 00, 0000}

\begin{abstract}
Accurate predictions for the non-linear matter power spectrum are needed to confront theory with observations in current and near future weak lensing and galaxy clustering surveys. We propose a computationally cheap method to create an emulator for modified gravity models by utilizing existing emulators for $\lcdm$. Using a suite of {\it N}-body simulations we construct a fitting function for the enhancement of both the linear and non-linear matter power spectrum in the commonly studied Hu-Sawicki $f(R)$ gravity model valid for wave-numbers $k \lesssim 5-10 \hMpc$ and redshifts $z \lesssim 3$. We show that the cosmology dependence of this enhancement is relatively weak so that our fit, using simulations coming from only one cosmology, can be used to get accurate predictions for other cosmological parameters. We also show that the cosmology dependence can, if needed, be included by using linear theory, approximate {\it N}-body simulations (such as COLA) and semi-analytical tools like the halo model. Our final fit can easily be combined with any emulator or semi-analytical models for the non-linear $\lcdm$ power spectrum to accurately, and quickly, produce a non-linear power spectrum for this particular modified gravity model. The method we use can be applied to fairly cheaply construct an emulator for other modified gravity models. As an application of our fitting formula we use it to compute Fisher-forecasts for how well galaxy clustering and weak lensing in a Euclid-like survey will be at constraining modifications of gravity.
\end{abstract}

\maketitle

\section{Introduction}\label{sec:intro}

One of the objectives of many next generation surveys such as Euclid \cite{2014IAUS..306..375S} and LSST \cite{2008arXiv0805.2366I} is to look for and constrain any deviations from the predictions of general relativity (GR). Modifications of gravity have been studied quite extensively over the last decade (see e.g. \cite{2016RPPh...79d6902K,2012PhR...513....1C} and references within). Such modifications, when they are in agreement with local and astrophysical tests of gravity, usually reduce to having most of their interesting effects in the non-linear regime of structure formation.

Testing such models, and extracting the maximum information that is contained in the data gathered from current and future galaxy and weak-lensing surveys, require us to include non-linear scales. This requires theoretical predictions for the matter power spectrum on these scales.

Currently this is either done using semi-analytical predictions and/or fits like \halofit \cite{2003MNRAS.341.1311S,2012ApJ...761..152T} and \hmcode \cite{2015MNRAS.454.1958M,2016MNRAS.459.1468M} (which is implemented in commonly used Boltzmann codes such as \camb \cite{Lewis:2002ah} and \class \cite{2011arXiv1104.2932L}) or using an emulator such as \cosmicemu \cite{2010ApJ...713.1322L,2014ApJ...780..111H,2016JCAP...08..008C} and \euclidemu \cite{2018arXiv180904695E}. An emulator is constructed by performing a large number of {\it N}-body simulations in the parameter-space and then performing an interpolation to obtain the power spectrum for any parameter combination of interest. This can be quite expensive to make, but once it's created it can provide non-linear spectra (usually from linear spectra) almost for free.

For $\lcdm$ both of the approaches above have been adopted and used to provide constraints from observations. For the case of modified gravity models the only fit provided so far is \mghalofit \cite{2014ApJS..211...23Z,2011JCAP...08..005H} which has a modified \halofit that was calibrated using {\it N}-body simulation data to the Hu-Sawicki model \cite{2007PhRvD..76f4004H}. For a coupled Dark Energy model, a fitting function was developed in \cite{Casas:2015qpa} and was used to forecast the constraints on the coupling parameter $\beta$. To be able to derive constraints, or to provide forecasts for how well future experiments will constrain deviations from GR, we need to have an accurate model for the non-linear matter power spectrum. This generally has to be derived on a model by model basis.

However very recently an interesting semi-analytical method based on the halo model and nonlinear perturbation theory was proposed in \cite{2018arXiv181205594C} and shown to be able to get $1-3\%$ accuracy out to $k\sim 1\hMpc$.

In this paper we will consider the Hu-Sawicki model as this is a representative model when it comes to using cosmology and astrophysics to constrain deviations from GR. Our aim is to provide the community with a precise fitting function for this model that can be used for example for making forecasts for future surveys. Instead of providing a fit for $P_{f(R)}$ directly, we present a fitting function for the enhancement $P_{f(R)} / P_{\lcdm}$ (both for the linear and non-linear power spectrum) as function of scale, redshift and the model parameter $f_{R0}$ which controls the size of the deviations from GR (GR is recovered as $f_{R0} \to 0$). This enhancement, as we will show, has a fairly weak cosmology dependence and we can therefore fit it using simulations from only one cosmology saving a lot of computational time. Our function have been fitted using a large suite of available {\it N}-body simulation data and can easily be incorporated in a Boltzmann code like \camb or \class to scale from a non-linear $P(k)$ for $\lcdm$ (created for example using an emulator like \euclidemu) to a non-linear $P(k)$ for the Hu-Sawicki model.

The requirements for the accuracy of the matter power spectrum is dictated by large upcoming surveys like LSST \cite{2008arXiv0805.2366I} and Euclid \cite{2014IAUS..306..375S}. Estimates for how accurate the power spectrum needs to be to take full advantage of the statistical power of such surveys varies from $1-2\%$ \cite{2005APh....23..369H} down as small as $0.5\%$ \cite{2012JCAP...04..034H} for scales $k\lesssim 10 \hMpc$. However this is ignoring\footnote{Also note that common {\it N}-body algorithms in state of the art codes disagree at the $\sim 1\%$ level already at $k = 1\hMpc$ and at the $\sim 3\%$ level at $k = 10 \hMpc$ \cite{2016JCAP...04..047S}.} model uncertainties in the way baryonic feedback affects the matter power spectrum. Baryonic effects are expected (from simulations and observations) to be as large as $10-30\%$ for scales $1\lesssim k \lesssim 10\hMpc$ \cite{2018MNRAS.480.3962C,2015JCAP...12..049S}. The accuracy of the newly released \euclidemu is quoted as being $\sim 1\%$ accurate for $k \lesssim 1\hMpc$ and for redshifts $z \lesssim 3.5$. Based on these considerations our aim is to produce a fit that is $\sim 1\%$ accurate for scales $k\lesssim 1\hMpc$ and $<5\%$ accurate for scales $1 < k < 10\hMpc$ and covering redshifts $z \lesssim 3.5$.

The setup of this paper is as follows: in \refSec{sec:sim} we describe the simulations we have used, in \refSec{sec:variation} we discuss the cosmology dependence of the enhancement $P_{f(R)} / P_{\lcdm}$, in \refSec{sec:fitfunc} we describe the fitting function we have created together with some tests, in \refSec{sec:forecasts} we show an application of the fitting formula by computing forecasts for how well galaxy clustering and weak-lensing in a Euclid-like survey will be at constraining $f(R)$ gravity before we conclude in \refSec{sec:conc}.

\section{Simulations}\label{sec:sim}

We take advantage of a large set of simulations to make the fit.\footnote{All the power spectrum data that we used are available at https://github.com/HAWinther/FofrFittingFunction} For more about how the modified gravity simulations are performed see e.g. \cite{2015MNRAS.454.4208W} and references within.

The main simulation suite we use is ELEPHANT \cite{2018MNRAS.476.3195C} (WMAP9 cosmology) which has $N=1024^3$ particles, $L = 1024 \Mpch$, $\Omega_m = 0.281$, $\Omega_b = 0.046$, $h = 0.6974$, $n_s = 0.971$ and $\sigma_8 = 0.820$ ($A_s = 2.3\cdot 10^{-9}$) with $|f_{R0}| = \{10^{-4},10^{-5},10^{-6},0\}$. These simulations were run with the \ecosmog code \cite{2012JCAP...01..051L}.

For the same cosmology as above we have also run extra simulations (also with the \ecosmog code) using $N = 256^3$ particles in a $L = 200 \Mpch$ box with $|f_{R0}| = \{10^{-5},5\cdot 10^{-6},2\cdot 10^{-5},5\cdot 10^{-5},0\}$. This simulation suite contain simulations of $\lcdm$ and $f(R)$ gravity with $|f_{R0}| = 10^{-5}$ for $\sigma_8 = 0.88$ and $0.72$ that allows us to test the $\sigma_8$ dependence of the modified gravity power spectrum enhancement.

We also use, mainly for testing and validation, the DUSTGRAIN simulations suite \cite{2018arXiv180604681G} (Planck 2015 cosmology) which has $N=768^3$ particles, $L = 750 \Mpch$, $\Omega_m = 0.31345$, $\Omega_b = 0.0481$, $h = 0.6731$, $n_s = 0.9658$ and $\sigma_8 = 0.842$ ($A_s = 2.199\cdot 10^{-9}$) with $|f_{R0}| = \{10^{-4},5\cdot 10^{-5}, 10^{-5},0\}$. This simulation suite contain simulations of $\lcdm$ and $f(R)$ gravity with $|f_{R0}| = 10^{-5}$ and with $\Omega_m = 0.2$ and $0.4$ that allows us to test the $\Omega_m$ dependence on the power spectrum. All these simulations have the same value of $\sigma_8$ and were run with the \mggadget code \cite{2013MNRAS.436..348P}.

All the $f(R)$ simulations mentioned above have corresponding $\lcdm$ simulations run with the same initial conditions that allows us to extract ratios $P_{f(R)} / P_{\lcdm}$ that (on the largest scales) are free of cosmic variance. For each simulation we have about $\sim 30$ redshifts between $z=0$ and $z=3$ that we use to compute the fit.

\newText{The power spectra used for the fitting functions have been estimated using \powmes \cite{2009MNRAS.393..511C} and (for DUSTGRAIN) by codes made by the authors. These codes have been tested and shown to give accurate results so we don't expect any bias due to different power-spectrum evaluation codes. What could give rise to a bias is that some of the data we use comes from different {\it N}-body codes, but as shown in \cite{2015MNRAS.454.4208W} even though the actual power spectrum varies between different code-types \cite{2016JCAP...04..047S} the enhancement $P_{f(R)}(k)/P_{\lcdm}(k)$ (with both spectra computed by the same code) does not and this is all that goes into our fit below.}

The effect of the size of the box, mass-resolution and cosmic variance on the matter power spectrum was investigated in \cite{2013MNRAS.428..743L}. We find that the spectra we extract can be trusted down to scales of $k\sim 5-10\hMpc$ depending on redshift.

\section{Variation of enhancement with cosmological parameters}\label{sec:variation}

The fit for $P_{f(R)}/P_{\lcdm}$ we perform in this paper is using {\it N}-body data from one single cosmology. The reason is that the cosmology dependence of this ratio is expected to be weak for almost all of the standard parameters with the possible exception of $\Omega_m$ and $\sigma_8$ as these correlate with the efficiency of screening and with the growth-rate of the matter density perturbation. There is also one additional effect that is potentially significant, which is how degenerate the enhancement is with baryonic feedback (see e.g. \cite{2013MNRAS.436..348P,2018arXiv180509824A} for a discussion on the size of these effects compared to the modified gravity enhancement).

In this section we will go through the different cosmological parameters and check how much the modified gravity enhancement changes. We will use linear perturbation theory, the halo model, fast approximative {\it N}-body simulations \cite{2017JCAP...08..006W} and full {\it N}-body simulations (for the cases we have this available) to investigate this.

By the halo model we mean the prediction of \cite{2014JCAP...03..021L}, which combines the modified linear power spectrum with a modified 1-halo contribution and a quasilinear correction motivated by higher-order perturbation theory \cite{2009PhRvD..79l3512K}. It incorporates the chameleon mechanism through an implementation of the thin-shell approximation in the spherical collapse model \cite{2013PhRvD..87l3511L}. This generates a mass and environment dependent spherical collapse density, from which an environmentally averaged modified peak threshold is determined that is used to compute the $f(R)$ modification and chameleon screening effects on the halo mass function and concentration determining the 1-halo contribution. A comparison to other modelling techniques of the modified nonlinear matter power spectrum in $f(R)$ gravity can be found in \cite{2014AnP...526..259L}.

The approximate {\it N}-body simulations we use is the COLA implementation of $f(R)$ gravity. COLA simulations are $\mathcal{O}(100-1000)$ faster than high-resolution {\it N}-body simulations, but can reproduce the enhancement of the power spectrum to \% -level accuracy down to fairly non-linear scales $k\lesssim 1-5 \hMpc$.

\subsubsection{Massive neutrinos}
\begin{figure*}
\includegraphics[width=0.99\columnwidth]{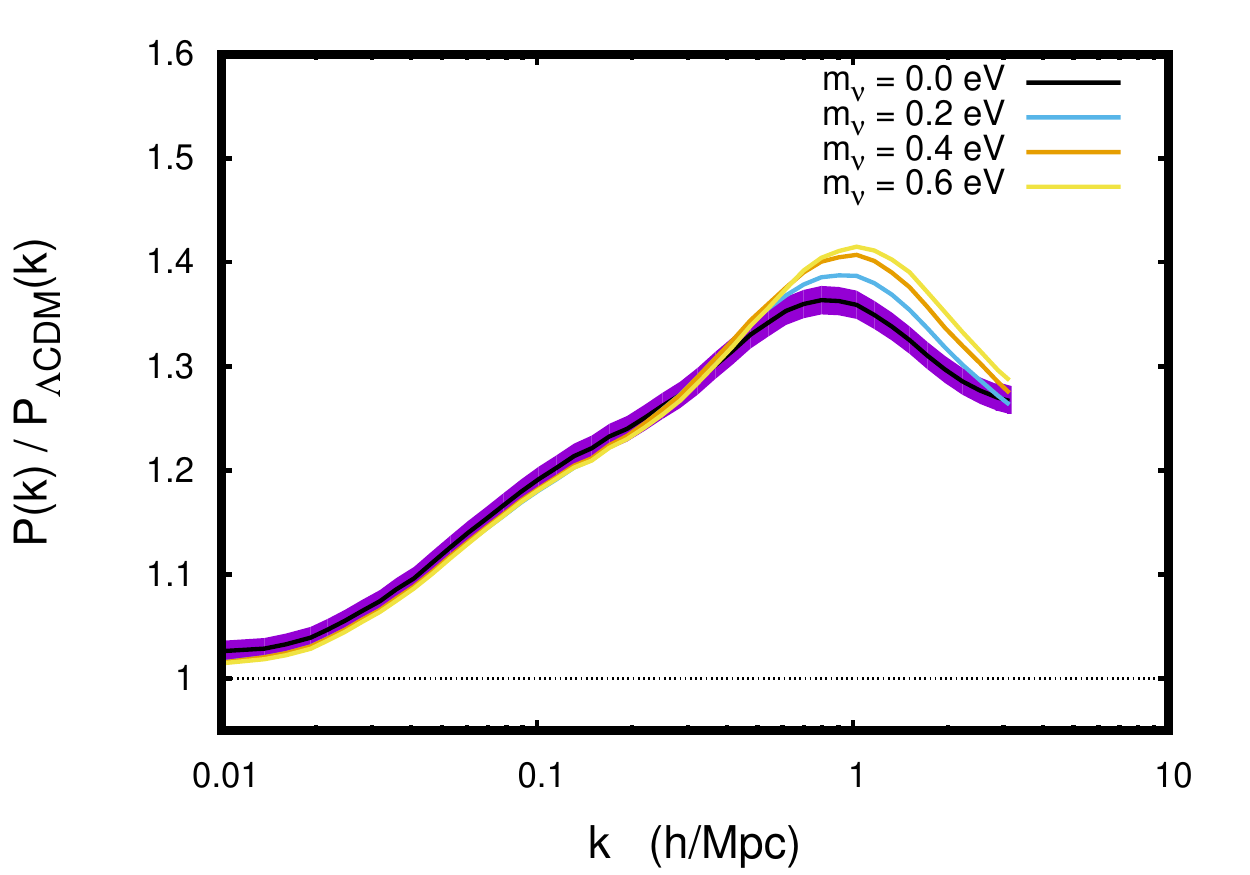}
\includegraphics[width=0.99\columnwidth]{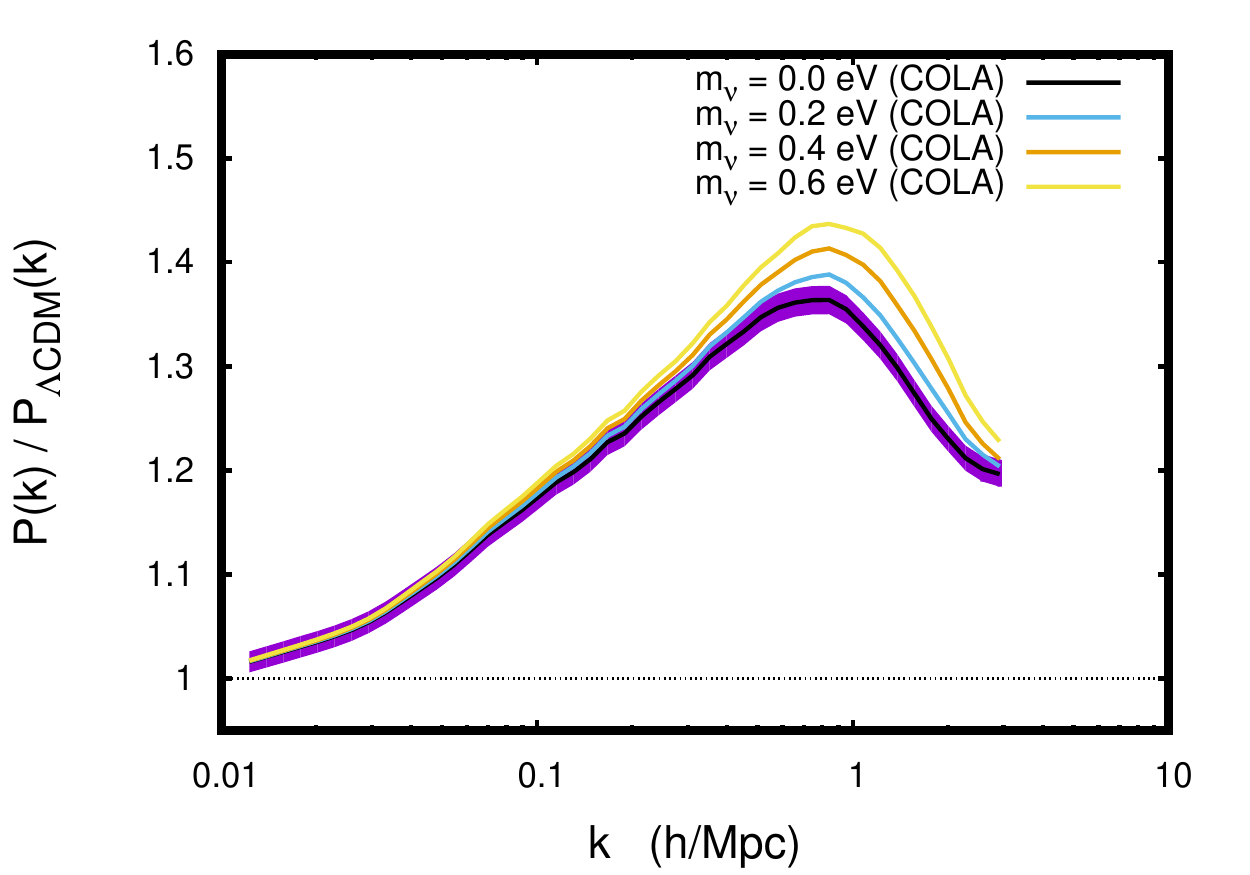}
  \caption{Estimation for the variation of the boost-factor $\frac{P(k)}{P_{\lcdm}(k)}$ with neutrino mass for $|f_{R0}| = 10^{-4}$ at $z=0$ based on {\it N}-body simulations (left) and COLA simulation (right). The shaded region correspond to $\pm$ 1\% of the central value. Note that the difference between the left and the right plot for $k\gtrsim 1 \hMpc$ comes from the fact that COLA simulations are not as accurate as high-resolution {\it N}-body simulations on very non-linear scales.}
\label{fig:nu}
\end{figure*}

The effects of massive neutrinos are highly degenerate with a modified gravity signal since massive neutrinos decrease the growth of structure on small scales, while modifications of gravity usualy enhances the growth. However we don't expect a big change in the enhancement (i.e. for fixed cosmological parameters) and this is what we see in \refFig{fig:nu} for {\it N}-body simulations with $|f_{R0}| = 10^{-4}$ \cite{2014MNRAS.440...75B}. The variation is seen to be at the sub-percent level for $k\lesssim 1\hMpc$ and $\sim 4\%$ around $k=1 \hMpc$ for the large value $m_\nu = 0.6$ eV.

\subsubsection{$\Omega_m$}
\begin{figure*}
\includegraphics[width=0.99\columnwidth]{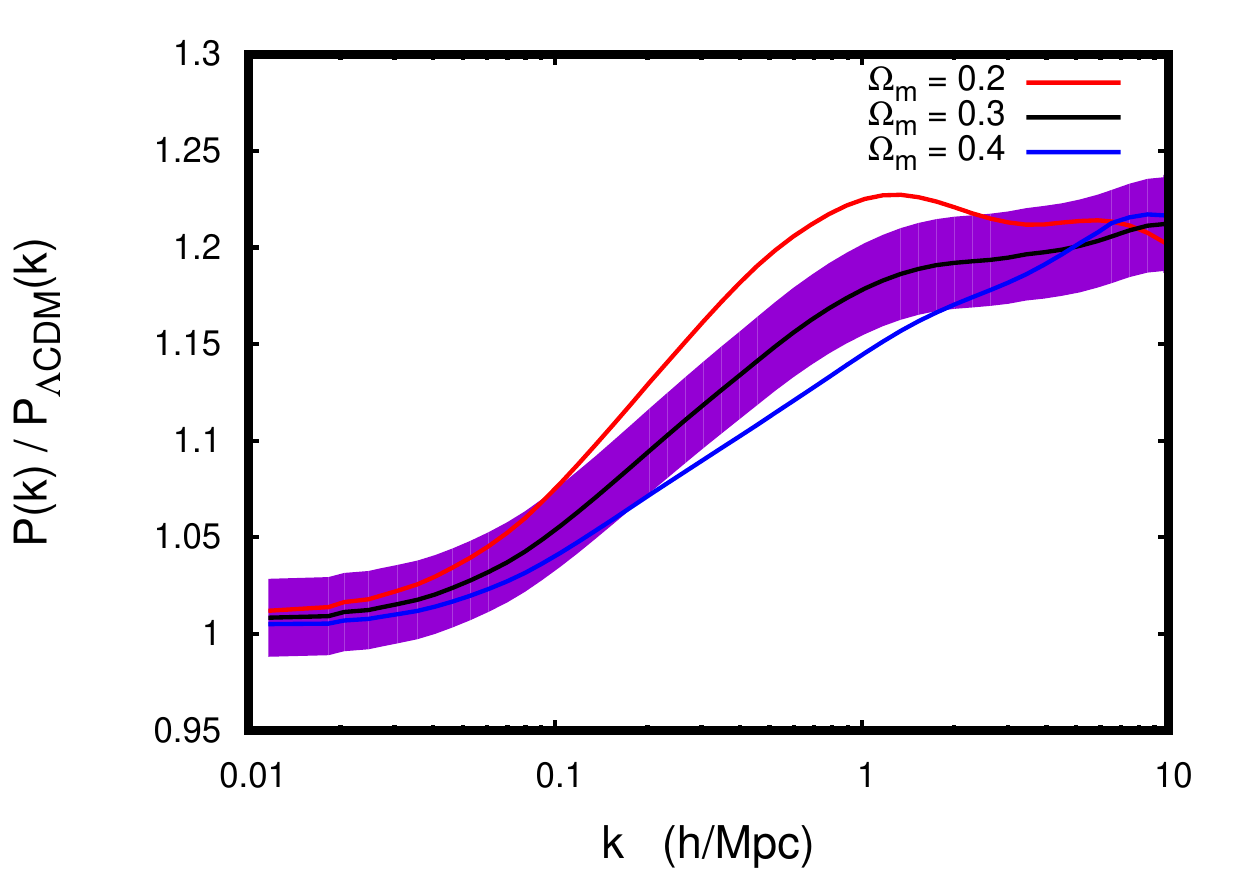}
\includegraphics[width=0.99\columnwidth]{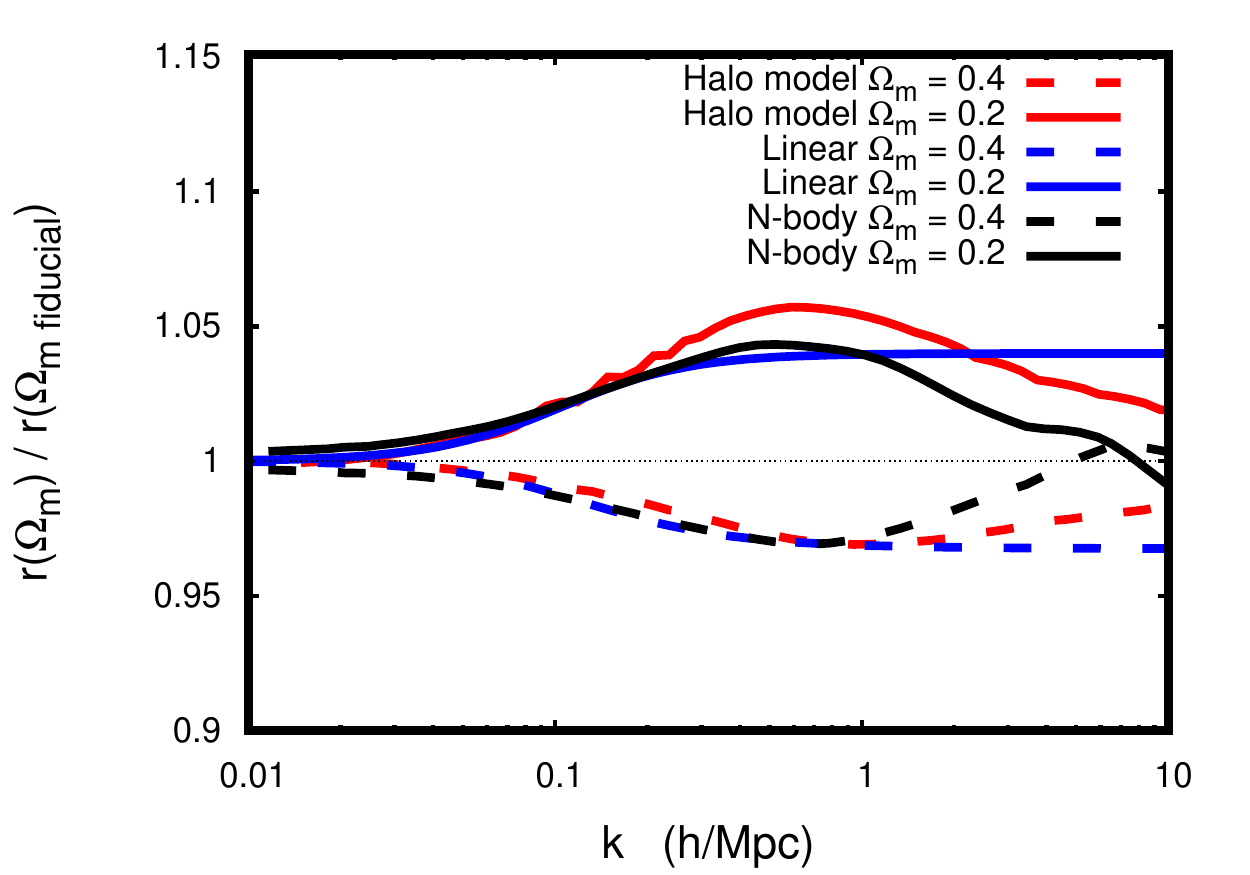}
\caption{Left: the enhancement $B = \frac{P(k)}{P_{\lcdm}(k)}$ for three different values of $\Omega_m$ for $|f_{R0}| = 10^{-5}$ at $z=0$ from {\it N}-body simulations. Right: variation of the enhancement $B(\Omega_m) / B(\Omega_m^{\rm fiducial})$ with respect to the fiducial value $\Omega_m^{\rm fiducial} = 0.3$ again for $|f_{R0}| = 10^{-5}$ at $z = 0$. The shaded region correspond to $\pm$ 2\% of the central value.}
\label{fig:omegamvariation}
\end{figure*}

Varying $\Omega_m$ changes the growth of linear perturbations so could potentially have a big effect on the enhancement. In \refFig{fig:omegamvariation} we show how much the enhancement changes for a fairly large variation of $\Omega_m$ (from $\Omega_m = 0.3$ to $\Omega_m = 0.2$ and $\Omega_m = 0.4$) based on {\it N}-body simulations. This figure also shows that the correction is accurately captured by linear theory and/or the halo model.

A simple fit to results computed using linear theory at $z=0$ \newText{shows that the enhancement of the power-spectrum $B_{\Omega_m \rm ~corr} \equiv P(k) /P_{\Omega_m = 0.3}(k)$} given by
\begin{align}
B_{\Omega_m \rm ~corr} \simeq 1 - a\frac{\Delta \Omega_m}{\Omega_m}\tanh\left(\frac{k}{k_*}\right)^b
\end{align}
\newText{with $\frac{\Delta \Omega_m}{\Omega_m} = \frac{\Omega_m - 0.3}{0.3}$, $a=0.105$, $b = 1.4$ and $k_* = 0.16(10^{-5}/|f_{R0}|)^{1/2} \hMpc$} (in general these parameters will depend on redshift). This gives us, for example, that a $10\%$ change in $\Omega_m$ leads only to a $\sim 1\%$ change in the enhancement so this is not a large effect, but it's straightforward to compute the correction using linear theory or the halo model if needed.

\subsubsection{Clustering amplitude}
\begin{figure*}
\includegraphics[width=0.99\columnwidth]{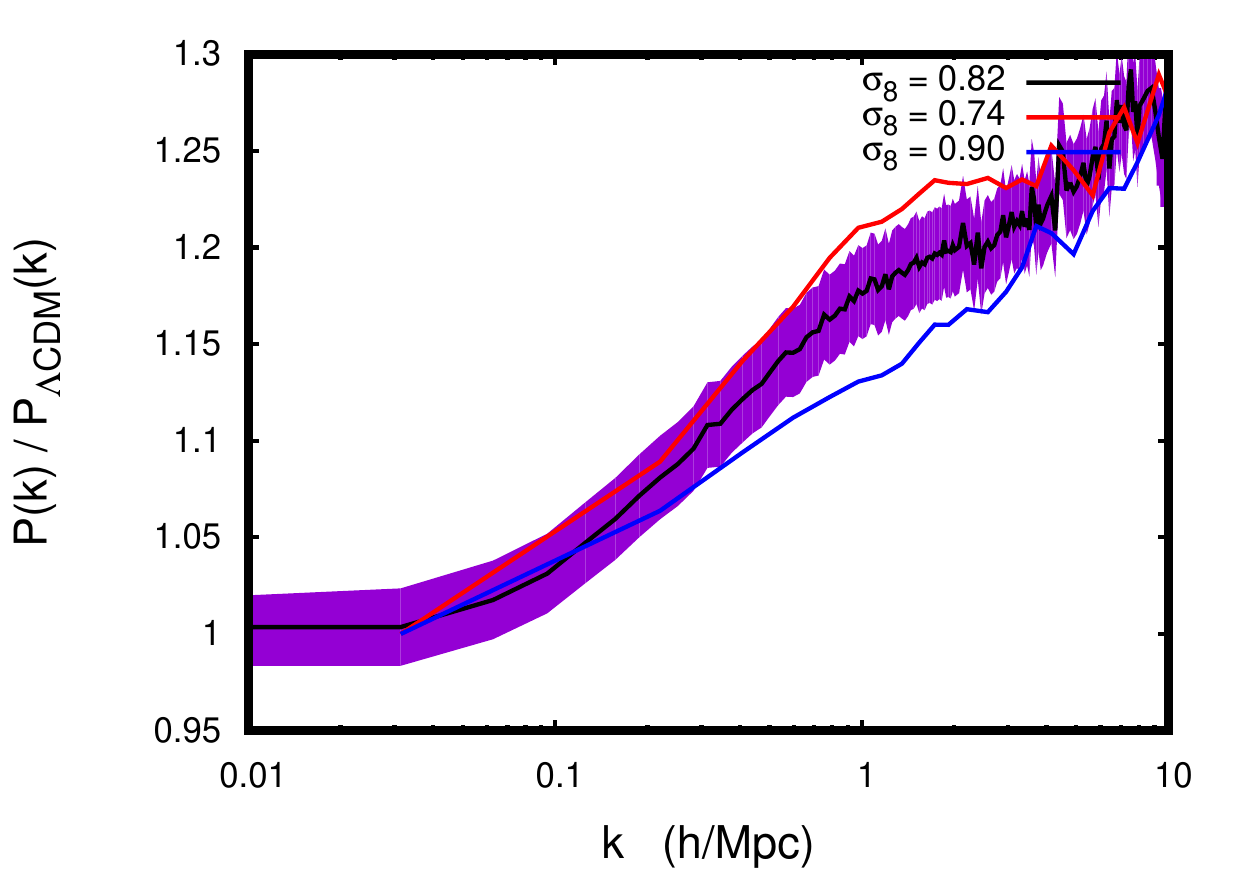}
\includegraphics[width=0.99\columnwidth]{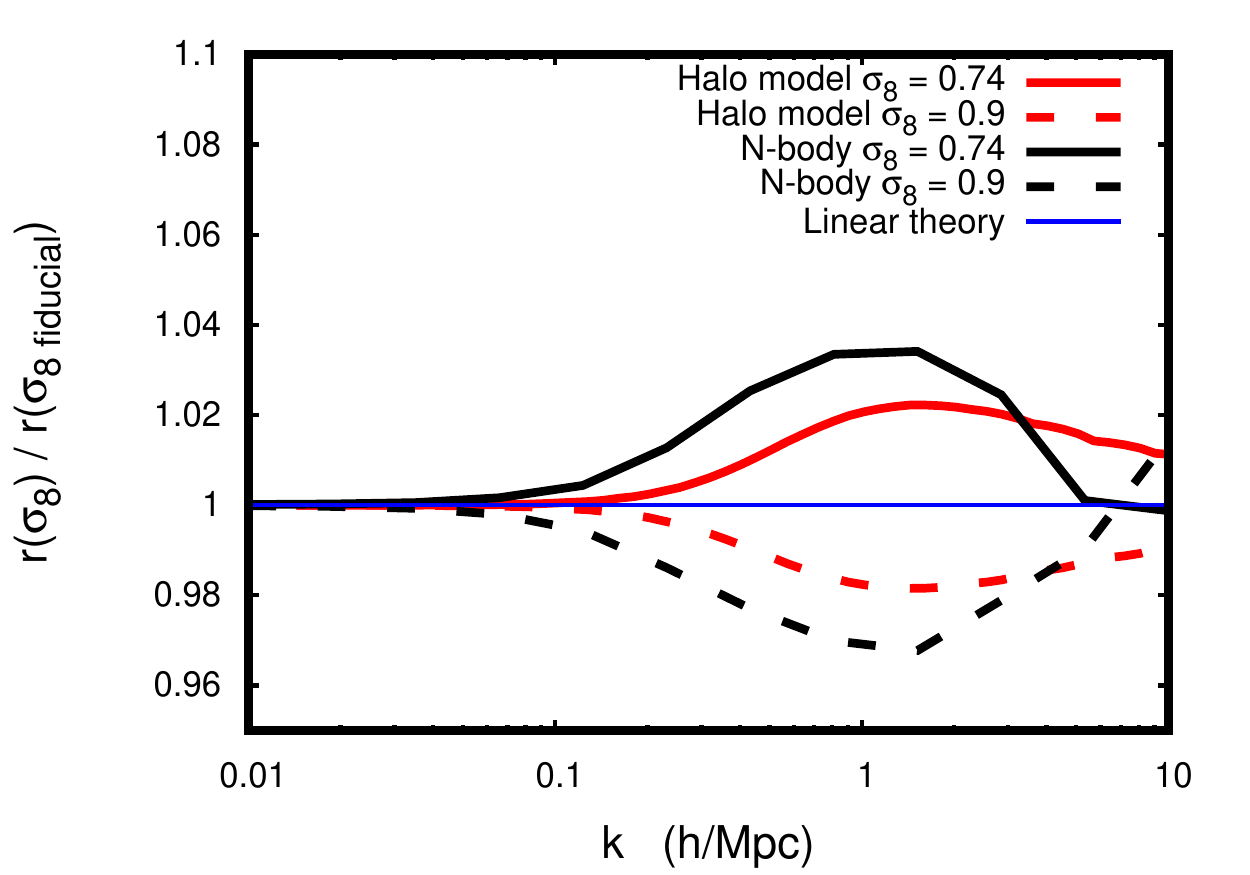}
\caption{Left: the enhancement $B = \frac{P(k)}{P_{\lcdm}(k)}$ for three different values of $\sigma_8$ for $|f_{R0}| = 10^{-5}$ at $z=0$ from {\it N}-body simulations. Right: variation of the enhancement $r(\sigma_8) / r(\sigma_8^{\rm fiducial})$ with respect to the fiducial value $\sigma_8^{\rm fiducial} = 0.82$ again for $|f_{R0}| = 10^{-5}$ at $z = 0$. The shaded region correspond to $\pm$ 2\% of the central value.}
\label{fig:sigma8variation}
\end{figure*}

In linear theory there is no variation with $\sigma_8$ (or more technically speaking with $A_s$, the primordial amplitude). However the amount of screening on non-linear scales depends crucially on how clustered matter is so this parameter could have a significant impact on non-linear scales. Even in $\lcdm$ an enhancement of $\sigma_8$ leads to a greater enhancement of the clustering on non-linear scales than as predicted by linear theory. In \refFig{fig:sigma8variation} we show how large this variation is based on {\it N}-body simulations together with predictions from both linear theory and the halo model.

A fit to our simulations with varying $\sigma_8$ ($0.72$, $0.8$ and $0.88$) shows that the effect of varying $\sigma_8$ can be described by a multiplicative correction to the enhancement which for $|f_{R0}| = 10^{-5}$ at $z=0$ and for $k\lesssim 10 \hMpc$ is approximately given by
\begin{align}
B_{\sigma_8 \rm ~corr} \simeq 1 + \frac{\Delta \sigma_8}{\sigma_8} \frac{k}{(1 + (k/k_*))^2}
\end{align}
where $k_* = 1.2 \hMpc$ and $\frac{\Delta \sigma_8}{\sigma_8} \equiv \frac{\sigma_8 - 0.8}{0.8}$. A $10\%$ deviation of $\sigma_8$ from it's fiducial value ($0.8$) leads to a $\sim 2-3\%$ level deviation in the enhancement for medium wave-numbers.

\subsubsection{Other parameters}
Linear perturbation theory predicts zero variation with other cosmological parameters such as the Hubble constant $H_0$ (when the fitting function is expressed in terms of $k$ in units of $\hMpc$) and the spectral index $n_s$. Changing the spectral index does modify the amplitude of clustering on small scales and could also influence screening, however we have checked using the halo model and COLA simulations that the expected variation is $\ll 1\%$ for all scales of interest within a reasonable variation in these parameters (here defined to be $3\sigma$ of the Planck 2018 cosmological constraints).

\section{Fitting functions}\label{sec:fitfunc}
\subsection{Fitting function for the linear enhancement}

\begin{figure}
\includegraphics[width=0.99\columnwidth]{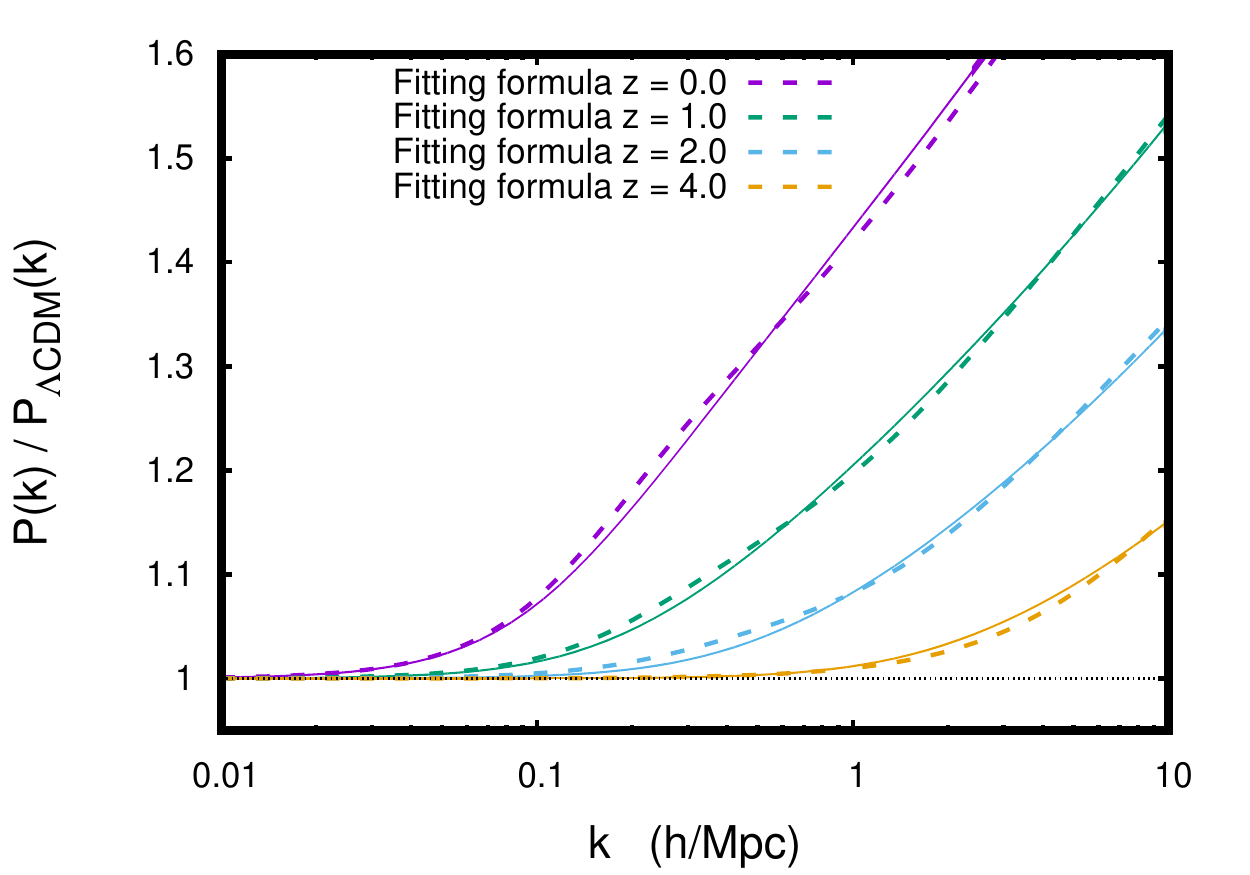}
\caption{Comparison of the fitting formula for the linear enhancement to the exact result from linear perturbation theory (here for $|f_{R0}| = 10^{-5}$).}
\label{fig:pofklincomp}
\end{figure}

At the level of linear perturbations the growth of matter perturbations are determined by
\begin{align}
\ddot{\delta} + 2H\dot{\delta} = \frac{3}{2}\Omega_m(a)H^2\delta\left(1 + \frac{1}{3}\frac{k^2}{k^2 + m^2a^2}\right)
\end{align}
where
\begin{align}
m^2(a) = \frac{H_0^2(\Omega_m + 4\Omega_\Lambda)}{2|f_{R0}|}\left(\frac{\Omega_m a^{-3} + 4\Omega_\Lambda}{\Omega_m + 4\Omega_\Lambda}\right)^3
\end{align}
from which it follows that the enhancement of the linear matter power spectrum for a general model is simply the enhancement of a fiducial model evaluated at $k^* = k (f_{R0} / f_{R0}^{\rm fid})^{1/2}$. We perform a fit using the fitting function
\begin{align}
\frac{P^{\rm linear~fid}(k,z)}{P^{\rm linear}_\lcdm(k,z)} &= 1 + \frac{(b(a)k)^2}{1 + c(a) k^2} + \nonumber\\
& + d(a)\left|\frac{\log(k)k}{k-1}\right|\arctan(e(a)k)
\end{align}
which is constructed to interpolate between the expected low and high $k$ limits ($\sim 1+bk^2$ for small $k$ and $\sim \log(k)$ for large $k$). 
Similar functional forms for the fitting functions of the enhancement within a coupled Dark Energy model were used in \cite{Casas:2015qpa}.
The functions, $X = b,c,d,e$, above are written as
\begin{align}
X(z,f_{R0}) = X_0(r) + X_1(r)(a-1) + X_2(r)(a-1)^2
\end{align}
\newText{where $r = \log(f_{R0} / f_{R0}^{\rm fid})$} leaving us with $12$ free parameters to fit. The best-fit from taking $|f_{R0}|^{\rm fid} = 10^{-5}$ using $\Omega_m = 0.281$ is shown in \refTable{table:fittingfunclinear} and in \refFig{fig:pofklincomp} we show a comparison to the true result.

\begin{table*}
\begin{tabular}{ | l | l | l | l | }
\hline
 & $i=0$ & $i=1$ & $i=2$ \\
\hline
$b_{0i}$ & 3.10     &  2.34466  & -1.86362 \\
\hline
$c_{0i}$ & 34.4951  &  28.8637  & -13.1302 \\
\hline
$d_{0i}$ & 0.14654  & -0.0100   & -0.14944 \\
\hline
$e_{0i}$ & 1.62807  &  0.71291  & -1.41003 \\
\hline
\end{tabular}
\label{table:fittingfunclinear}
\caption{The best-fit parameters for the enhancement of the linear power spectrum for $|f_{R0}^{\rm fid}| = 10^{-5}$.}
\end{table*}

\subsection{Fitting function for the non-linear enhancement}
The fitting function we use for our fit is given by
\begin{align}
\frac{P_{f(R)}(k,z)}{P_{\lcdm}(k,z)} = 1 + b(z,f_{R0})\frac{(1+ c(z,f_{R0}) \cdot k )}{(1+ d(z,f_{R0}) \cdot k )}\times\nonumber\\\ \times\, \arctan\left( e(z,f_{R0}) \cdot k)\right)^{f(z,f_{R0}) + g(z,f_{R0})\cdot k}
\end{align}
where (for $X = b,c,d,e,f,g$)
\begin{align}
X(z,f_{R0}) = X_0(r) + X_1(r)(a-1) + X_2(r)(a-1)^2
\end{align}
with
\begin{align}
X_i(r) = X_{i0} + X_{i1} r + X_{i2}r^2
\end{align}
where $r = \log(f_{R0} / f_{R0}^{\rm fid})$. This makes $54$ free parameters for the full $f_{R0}$, scale and redshift dependence.

\begin{table*}
\begin{tabular}{ | l | l | l | l | }
\hline
 & $i=0$ & $i=1$ & $i=2$ \\

\hline
$b_{0i}$ &    0.76878 &    0.22638 &    0.00759 \\
\hline
$b_{1i}$ &   -0.40537 &   -0.10711 &   -0.00102 \\
\hline
$b_{2i}$ &    0.00752 &    0.04846 &    0.01180 \\
\hline
$c_{0i}$ &    0.02886 &   -0.02438 &   -0.02963 \\
\hline
$c_{1i}$ &   -0.06382 &   -0.05196 &    0.02597 \\
\hline
$c_{2i}$ &   -0.40121 &   -0.03518 &    0.07688 \\
\hline
$d_{0i}$ &    1.00000 &    0.10901 &    0.12027 \\
\hline
$d_{1i}$ &    0.00000 &    0.08189 &    0.02492 \\
\hline
$d_{2i}$ &    0.00000 &   -0.05682 &   -0.02985 \\
\hline
$e_{0i}$ &    0.36951 &    0.14719 &    0.03127 \\
\hline
$e_{1i}$ &    0.10939 &    0.06176 &    0.02933 \\
\hline
$e_{2i}$ &   -0.34209 &   -0.13138 &   -0.01419 \\
\hline
$f_{0i}$ &    1.03544 &   -0.13912 &   -0.05656 \\
\hline
$f_{1i}$ &   -0.26277 &   -0.13231 &   -0.03389 \\
\hline
$f_{2i}$ &    0.23028 &    0.13132 &    0.07127 \\
\hline
$g_{0i}$ &    0.20246 &    0.06323 &    0.05229 \\
\hline
$g_{1i}$ &   -0.11611 &    0.06943 &    0.07807 \\
\hline
$g_{2i}$ &    0.10245 &    0.00296 &   -0.09770 \\
\hline

\end{tabular}
\begin{tabular}{ | l | l | l | l | }
\hline
 & $i=0$ & $i=1$ & $i=2$ \\

\hline
$b_{0i}$ &    0.93650 &   -0.03999 &    0.24007 \\
\hline
$b_{1i}$ &   -0.54583 &    0.30370 &    0.18820 \\
\hline
$b_{2i}$ &    0.63480 &    0.36096 &    0.66583 \\
\hline
$c_{0i}$ &   -0.02906 &    0.00062 &    0.01222 \\
\hline
$c_{1i}$ &   -0.09544 &   -0.00942 &   -0.03434 \\
\hline
$c_{2i}$ &   -0.34249 &   -0.01813 &   -0.05204 \\
\hline
$d_{0i}$ &    1.00000 &    0.39355 &    0.77661 \\
\hline
$d_{1i}$ &    0.00000 &    0.29088 &    0.47078 \\
\hline
$d_{2i}$ &    0.00000 &   -0.41149 &   -0.68192 \\
\hline
$e_{0i}$ &    0.49107 &    0.37630 &    0.26101 \\
\hline
$e_{1i}$ &    0.29782 &    0.48636 &    0.52563 \\
\hline
$e_{2i}$ &   -0.28714 &    0.03494 &    0.26626 \\
\hline
$f_{0i}$ &    0.92041 &   -0.09308 &    0.27038 \\
\hline
$f_{1i}$ &   -0.28239 &    0.07838 &    0.37029 \\
\hline
$f_{2i}$ &    0.53954 &    0.19496 &    0.19486 \\
\hline
$g_{0i}$ &    0.31864 &    0.03340 &   -0.00276 \\
\hline
$g_{1i}$ &    0.04570 &    0.07630 &    0.04616 \\
\hline
$g_{2i}$ &    0.13924 &   -0.00010 &    0.18990 \\
\hline

\end{tabular}
\begin{tabular}{ | l | l | l | l | }
\hline
 & $i=0$ & $i=1$ & $i=2$ \\

\hline
$b_{0i}$ &    0.57248 &    0.49880 &    0.57426 \\
\hline
$b_{1i}$ &    0.25469 &    0.36089 &   -0.30799 \\
\hline
$b_{2i}$ &    1.21637 &    0.07034 &    0.83164 \\
\hline
$c_{0i}$ &    0.00046 &    0.02574 &   -0.00936 \\
\hline
$c_{1i}$ &   -0.09012 &    0.01689 &    0.00221 \\
\hline
$c_{2i}$ &   -0.35585 &   -0.03070 &    0.00768 \\
\hline
$d_{0i}$ &    1.00000 &    0.77903 &    1.26756 \\
\hline
$d_{1i}$ &    0.00000 &    0.91964 &    1.44477 \\
\hline
$d_{2i}$ &    0.00000 &   -0.93633 &   -1.44129 \\
\hline
$e_{0i}$ &    2.31154 &   -0.20699 &   -0.65038 \\
\hline
$e_{1i}$ &    2.29822 &    0.26608 &    0.01792 \\
\hline
$e_{2i}$ &   -0.48319 &    0.60336 &    0.92704 \\
\hline
$f_{0i}$ &    1.21959 &   -0.25171 &   -0.14644 \\
\hline
$f_{1i}$ &    0.35388 &    0.12487 &    0.02003 \\
\hline
$f_{2i}$ &    1.02533 &    0.34599 &    0.08923 \\
\hline
$g_{0i}$ &    0.28475 &   -0.04719 &    0.09592 \\
\hline
$g_{1i}$ &   -0.15829 &    0.13977 &    0.36819 \\
\hline
$g_{2i}$ &    0.54118 &   -0.13489 &   -0.15783 \\
\hline

\end{tabular}
\caption{The best fit values using the data from $|f_{R0}| = \{10^{-5},5\cdot 10^{-6},10^{-6},10^{-7}\}$ with $|f_{R0}|^{\rm fid} = 5\cdot 10^{-6}$ (left), from $|f_{R0}| = \{5\cdot 10^{-5},10^{-5},5\cdot 10^{-6}\}$ with $|f_{R0}|^{\rm fid} = 10^{-5}$ (middle) and from $|f_{R0}| = \{10^{-4},5\cdot 10^{-5},10^{-5}\}$ with $|f_{R0}|^{\rm fid} = 5\cdot 10^{-5}$ (right).}
\label{table:results}
\end{table*}

\begin{figure*}
\includegraphics[width=0.99\columnwidth]{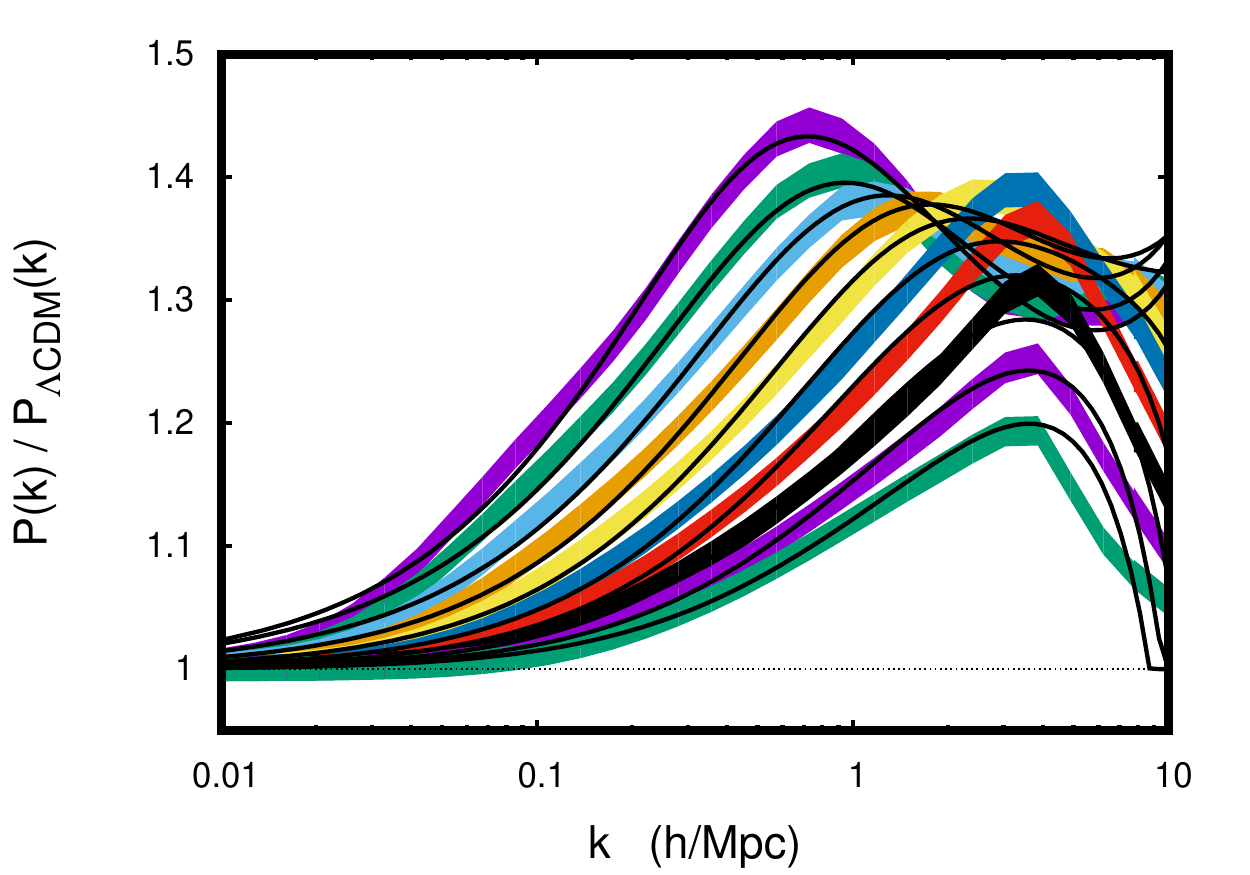}\includegraphics[width=0.99\columnwidth]{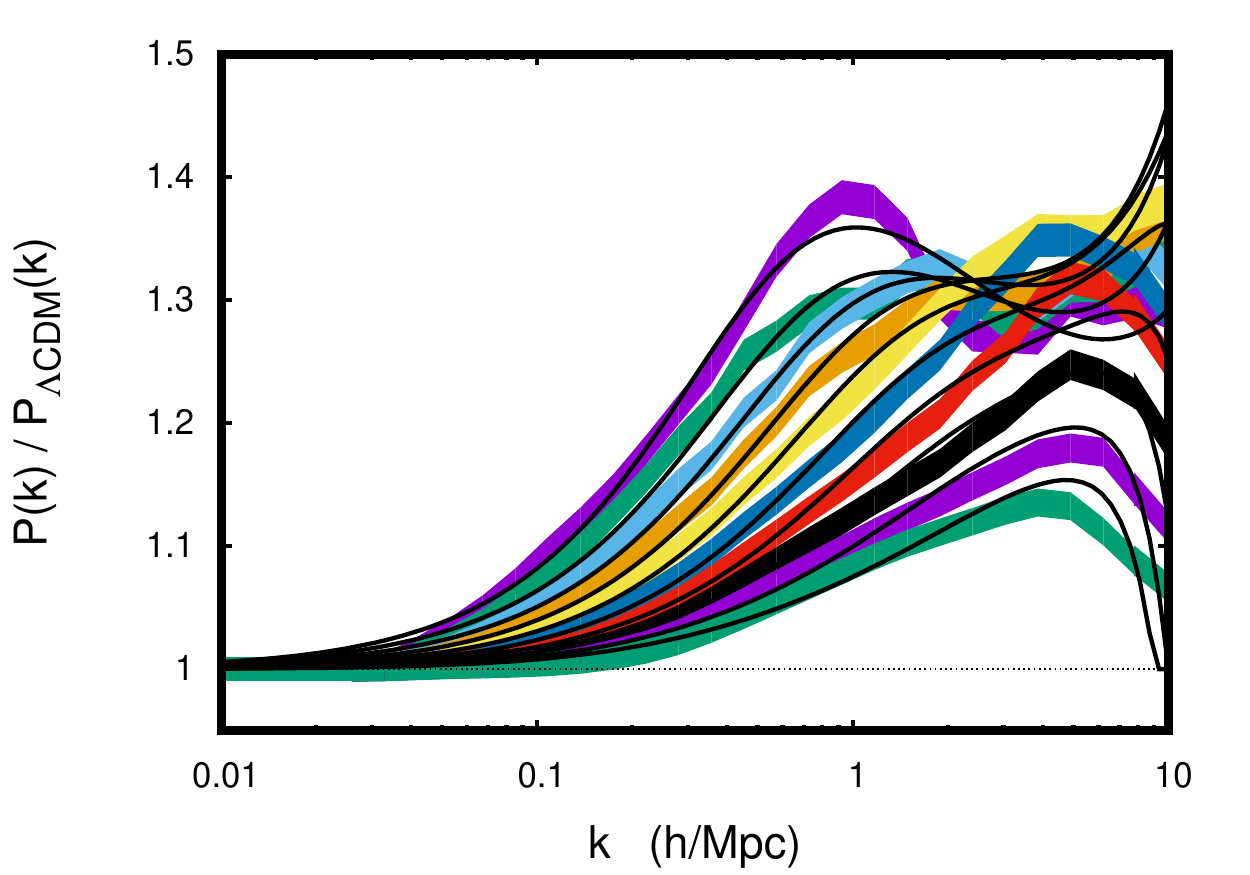}\\
\includegraphics[width=0.99\columnwidth]{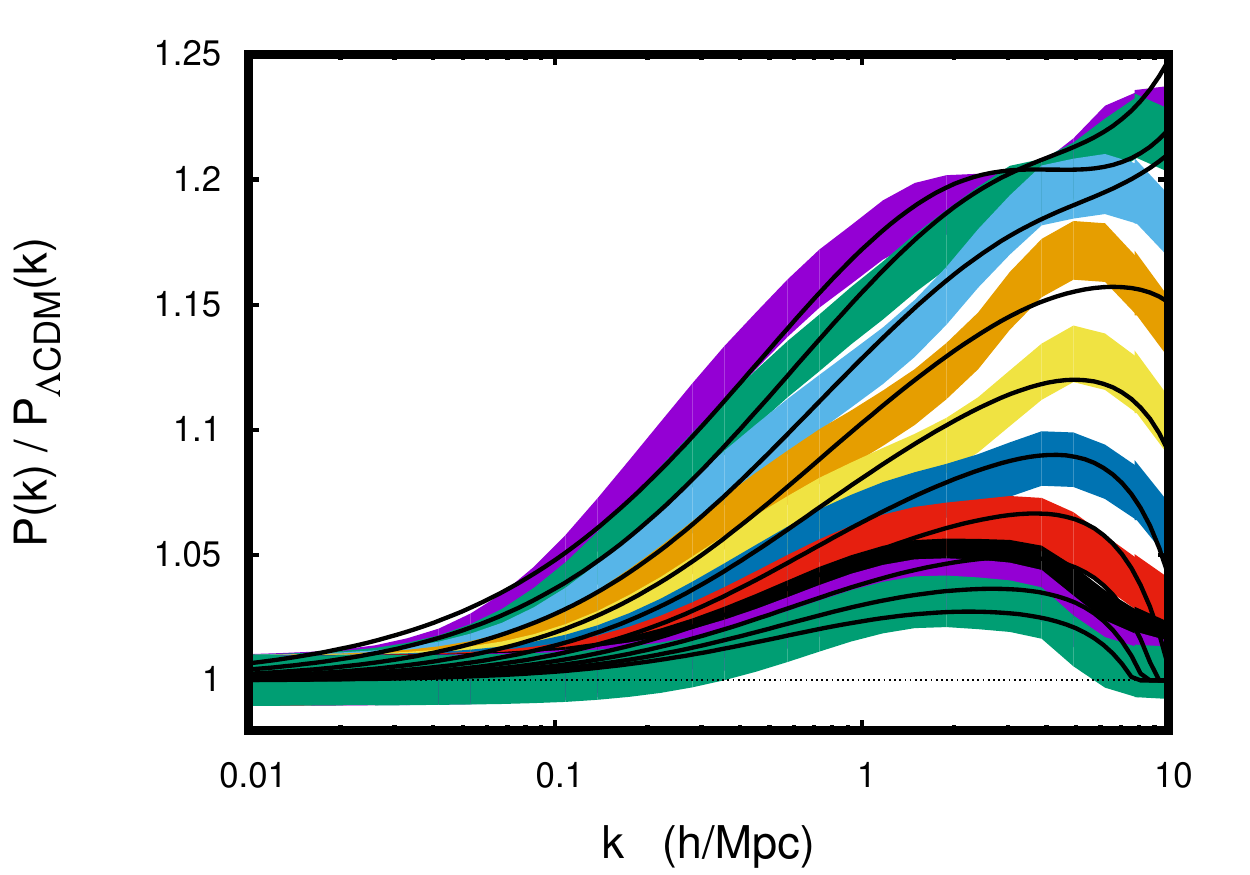}\includegraphics[width=0.99\columnwidth]{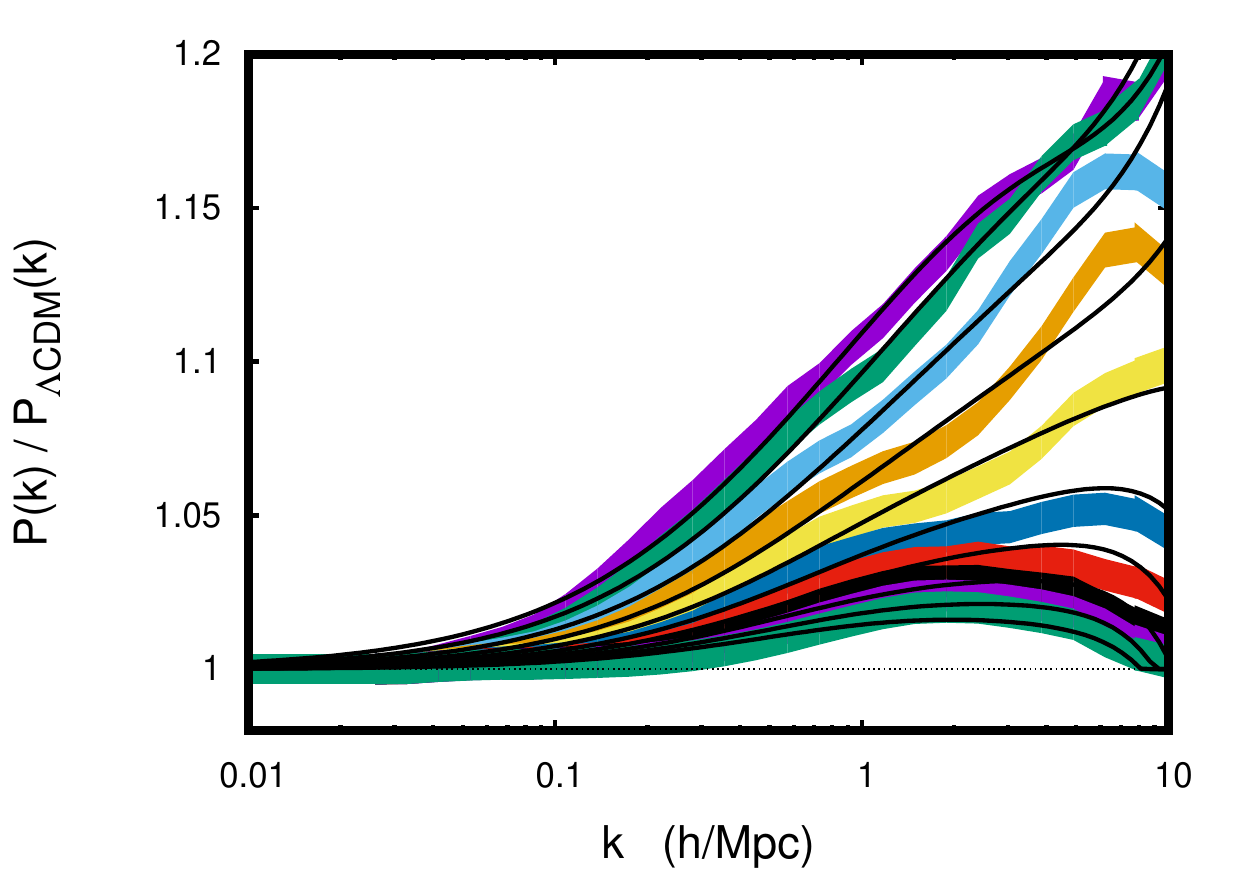}\\
\includegraphics[width=0.99\columnwidth]{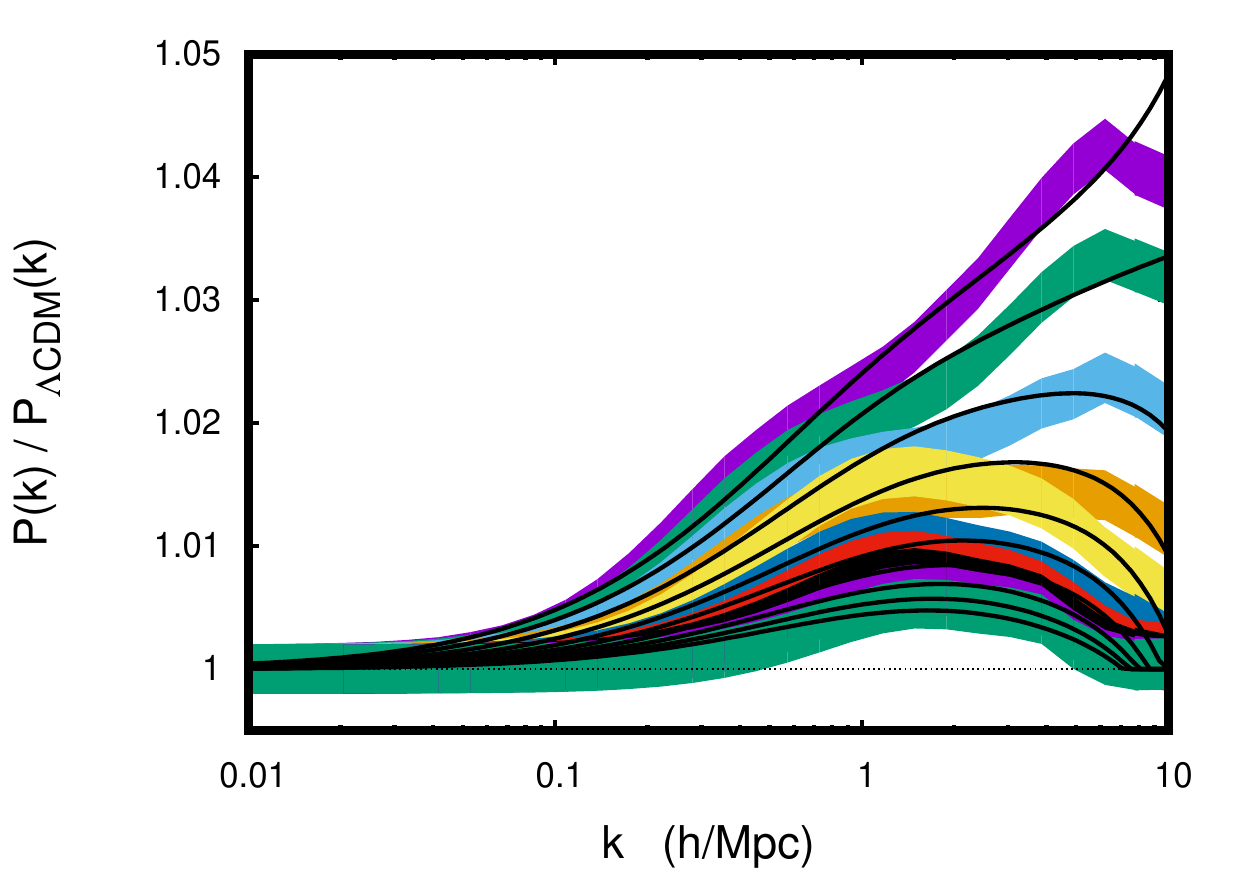}
\caption{Fitting function compared to {\it N}-body data for $|f_{R0}| = 10^{-4}$ (up left),  $|f_{R0}| = 5\cdot 10^{-5}$ (up right),  $|f_{R0}| = 10^{-5}$ (middle left),  $|f_{R0}| = 5\cdot 10^{-6}$ (middle right) and  $|f_{R0}| = 10^{-6}$ (bottom) and $z = \{0.0, 0.259, 0.518, 0.777, 1.036, 1.294, 1.554, 1.812, 2.071, 2.330\}$. The shaded region is to give the reader an idea of how good the fitting function is and denotes $\pm$ 1\% for all plots except for $|f_{R0}| = 5\cdot 10^{-6}$ and $|f_{R0}| = 10^{-6}$ where it's $\pm 0.5\%$ and $\pm 0.2\%$ respectively.}
\label{fig:fitvsnbody}
\end{figure*}

We choose to make three different fits: one using $|f_{R0}| = \{10^{-4},5\cdot 10^{-5},10^{-5}\}$ (high), one using $|f_{R0}| = \{10^{-5},5\cdot 10^{-6},10^{-6}\}$ (medium) and one using $|f_{R0}| = \{10^{-5},5\cdot 10^{-6},10^{-6},10^{-7}\}$ (low) and then interpolate between these (overlapping) fits. The best-fit parameters we find are given in \refTable{table:results} and the agreement as a function of scale and redshift can be seen in \refFig{fig:fitvsnbody}. This figure shows that the fit is good to $\lesssim 1$\% for most scales and redshifts, with the exception of the smallest scales for the largest values of $|f_{R0}|$.

\begin{figure}
\includegraphics[width=0.99\columnwidth]{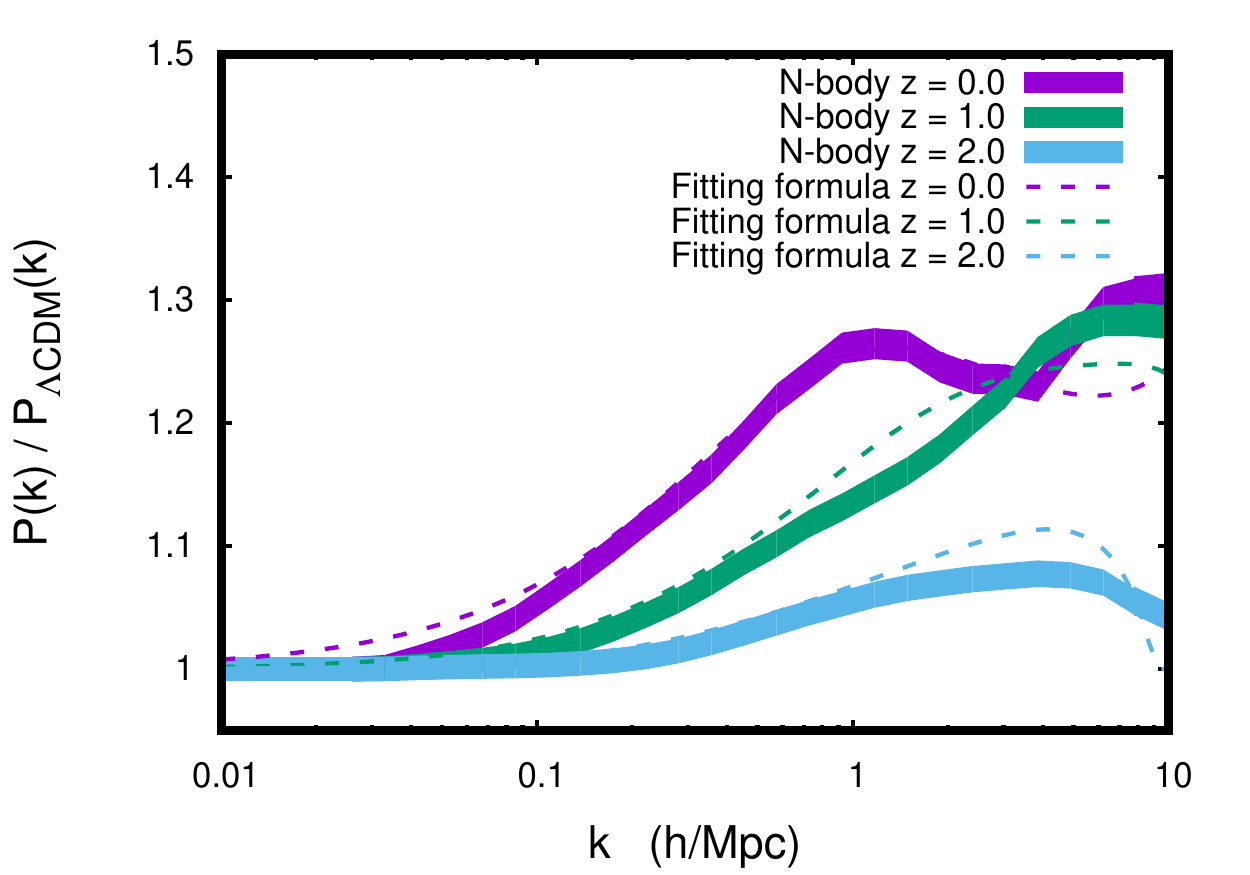}
\caption{Test of the fitting function to {\it N}-body data not used in the fit (left), i.e. simulations with $|f_{R0}| = 2\cdot 10^{-5}$. The shaded region correspond to $\pm$ 1\% of the central value.}
\label{fig:test}
\end{figure}

In \refFig{fig:test} we perform a test of our fitting function by predicting the enhancement for $|f_{R0}| = 2\cdot 10^{-5}$, a value that was not used to generate the fit. This value is in the middle of $|f_{R0}| = 10^{-5}$ and $|f_{R0}| = 5\cdot 10^{-5}$ and should therefor give us a good estimate for the accuracy of our fit. \newText{The agreement is within $\sim 1-2$\% for $k \lesssim 1\hMpc$ and $\sim 2-3$\% for $k\lesssim 10\hMpc$.}

\begin{figure}
\includegraphics[width=0.99\columnwidth]{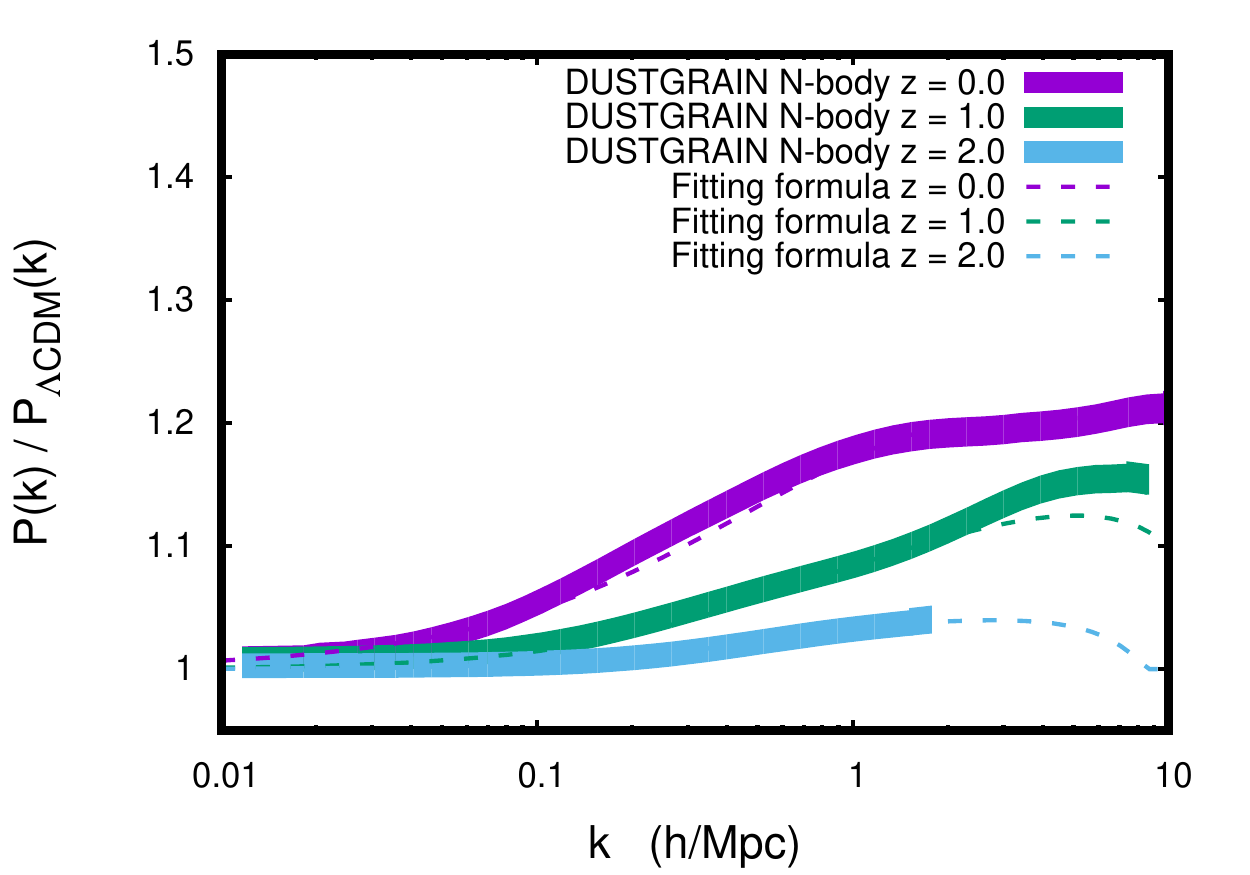}
\caption{Test of the fitting function to {\it N}-body data with a different background cosmology (DUSTGRAIN vs ELEPHANT) for $|f_{R0}| = 10^{-5}$. The shaded region correspond to $\pm$ 1\% of the central value.}
\label{fig:test_dustgrain}
\end{figure}

In \refFig{fig:test_dustgrain} we test our fitting function by comparing it to simulations with a different background cosmology from that is used to make the fit. As expected from the discussion in the previous section the agreement is very good.

These numbers should be contrasted to the enhancement of the matter power spectrum itself relative to $\lcdm$ which is typically $10-40\%$.

\begin{figure*}
\includegraphics[width=0.99\columnwidth]{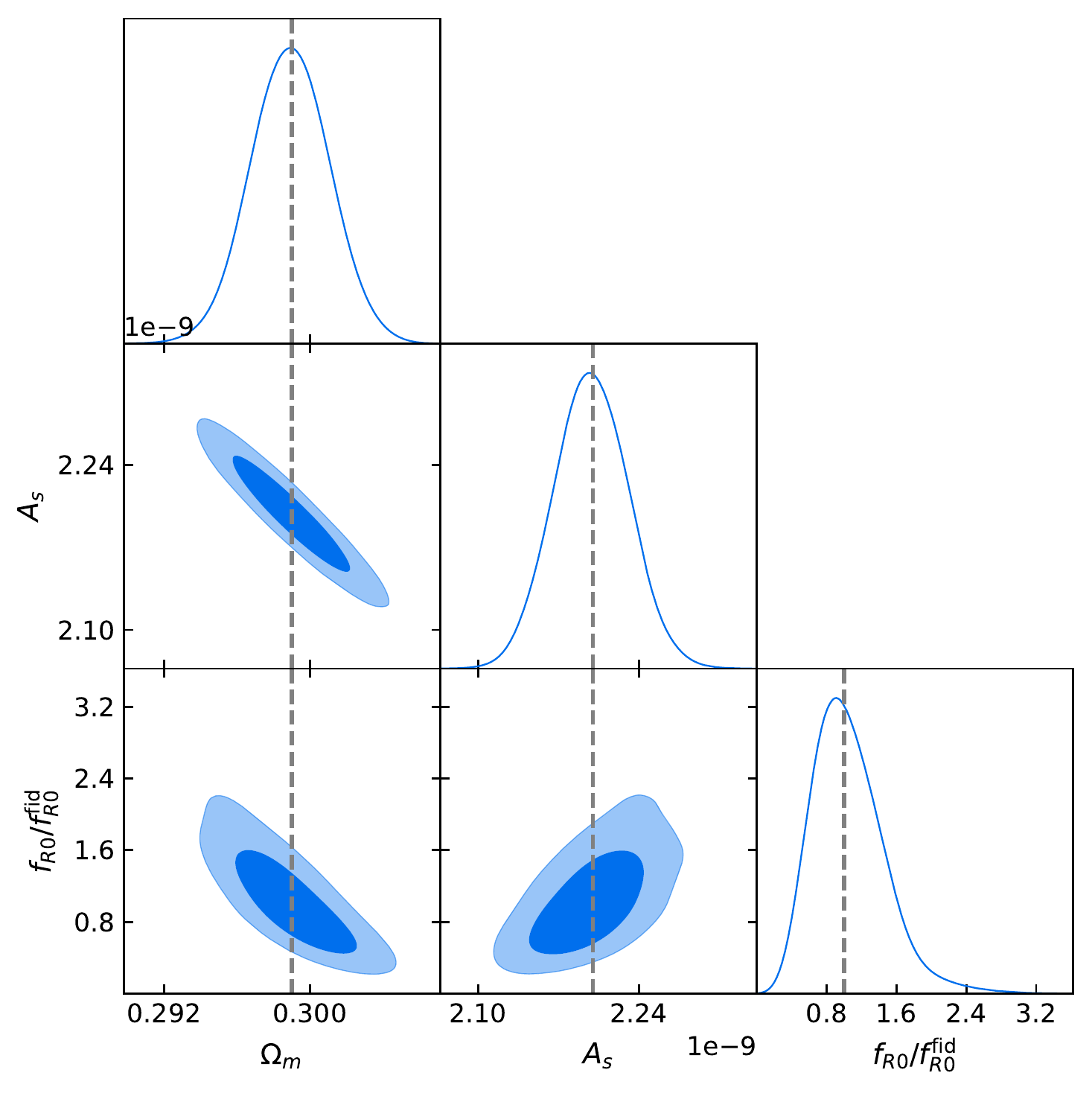}
\includegraphics[width=0.99\columnwidth]{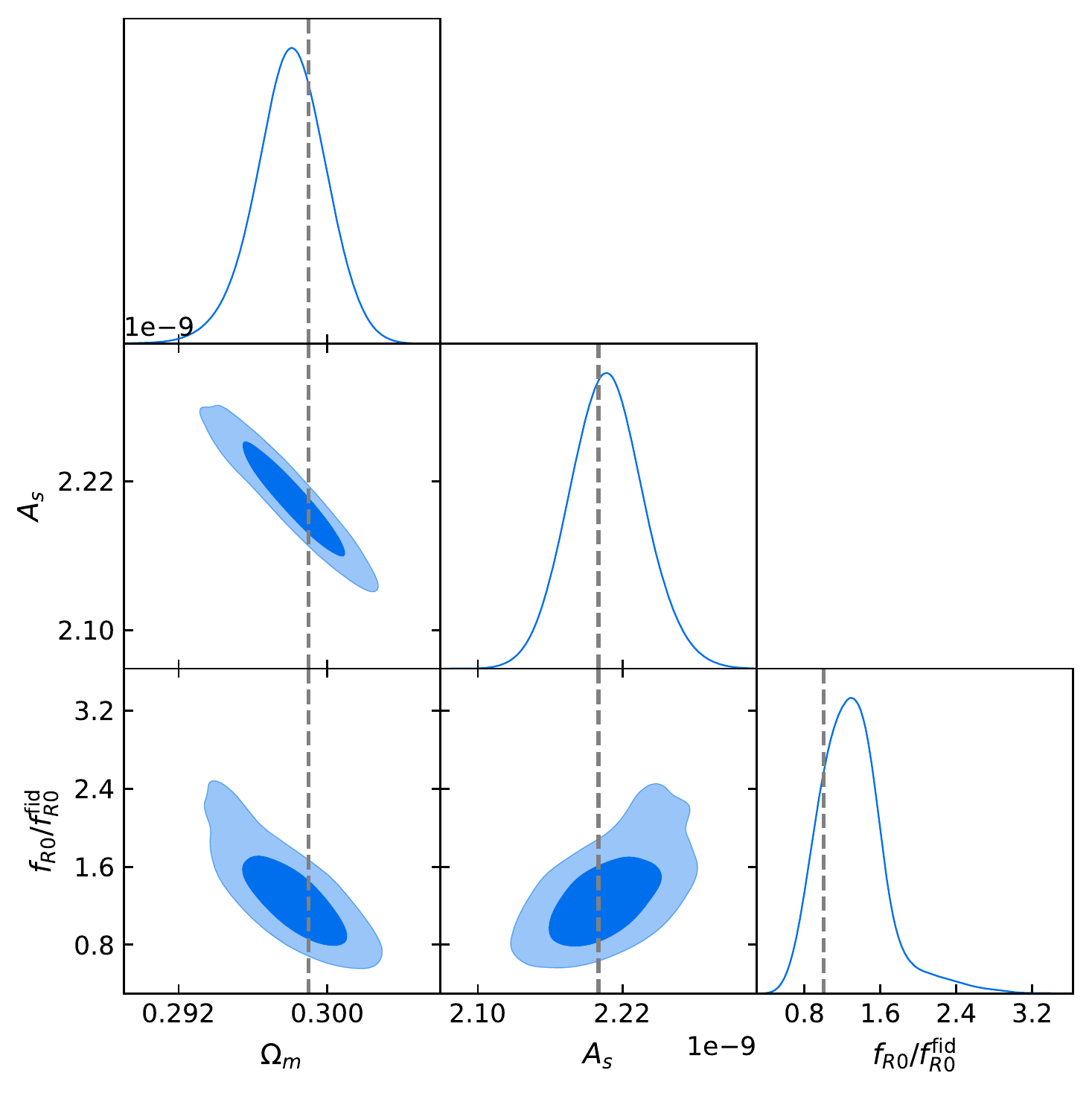}
\caption{Test of the fitting function by fitting to mock data, using the {\it N}-body enhancement for $|f_{R0}| = 2\cdot 10^{-5}$ around $z=1$. We show the fits for the two values $k_{\rm max } = 1 \hMpc $ (left) and $k_{\rm max} = 5 \hMpc $ (right).}
\label{fig:fitmockdata}
\end{figure*}

To test the fitting function we create mock data and try to perform a fit to $P(k,z)$ for a Euclid-like survey with $V = 50(\text{Gpc}/h)^3$ with $n_{\rm gal} = 10^{-3}(\Mpch)^{-3}$ using the diagonal likelihood
\begin{align}
  \log\mathcal{L} = - \frac{1}{2}\sum_{k,z} (P(k,z) - P^{\rm fid}(k,z))C^{-1}(P(k,z) - P^{\rm fid}(k,z))
\end{align}
where $C^{-1} = \frac{Vk^2\Delta k}{4\pi^2 (P^{\rm fid}(k,z) + 1/n)^2}$. \newText{The sum is over $6$ evenly spaced $z$-bins from $z=0$ to $z=2$ and $30$ logarithmically spaced $k$-bins between $k=10^{-4}\hMpc$ and $k=5\hMpc$.} This assumes Gaussian fluctuations on all scales which significantly underestimate the errors on non-linear scales and also does not take into account uncertainty of unknown baryonic physics. Thus any fit based on this would be completely dominated by the smallest scales. To get more realistic errors we try to take this into account by imposing a minimum $1\%$ error on non-linear scales starting at $k=0.5\hMpc$ and growing to $10\%$ at $k=10 \hMpc$. The fiducial power spectrum is generated using the Eisenstein-Hu fitting function for $\lcdm$ \cite{1998ApJ...496..605E,1999ApJ...511....5E}, converted to a non-linear power spectrum using \halofit and finally multiplied by the $f(R)$ enhancement found in simulations with $|f_{R0}| = 2\cdot 10^{-5}$. The result of fits to $P(k,z)$ around $z=1$ (where our fit is seen to deviate a bit from the {\it N}-body result) can be found in \refFig{fig:fitmockdata}. We also performed this test for other redshifts with similar results.

\begin{figure*}
\includegraphics[width=0.99\columnwidth]{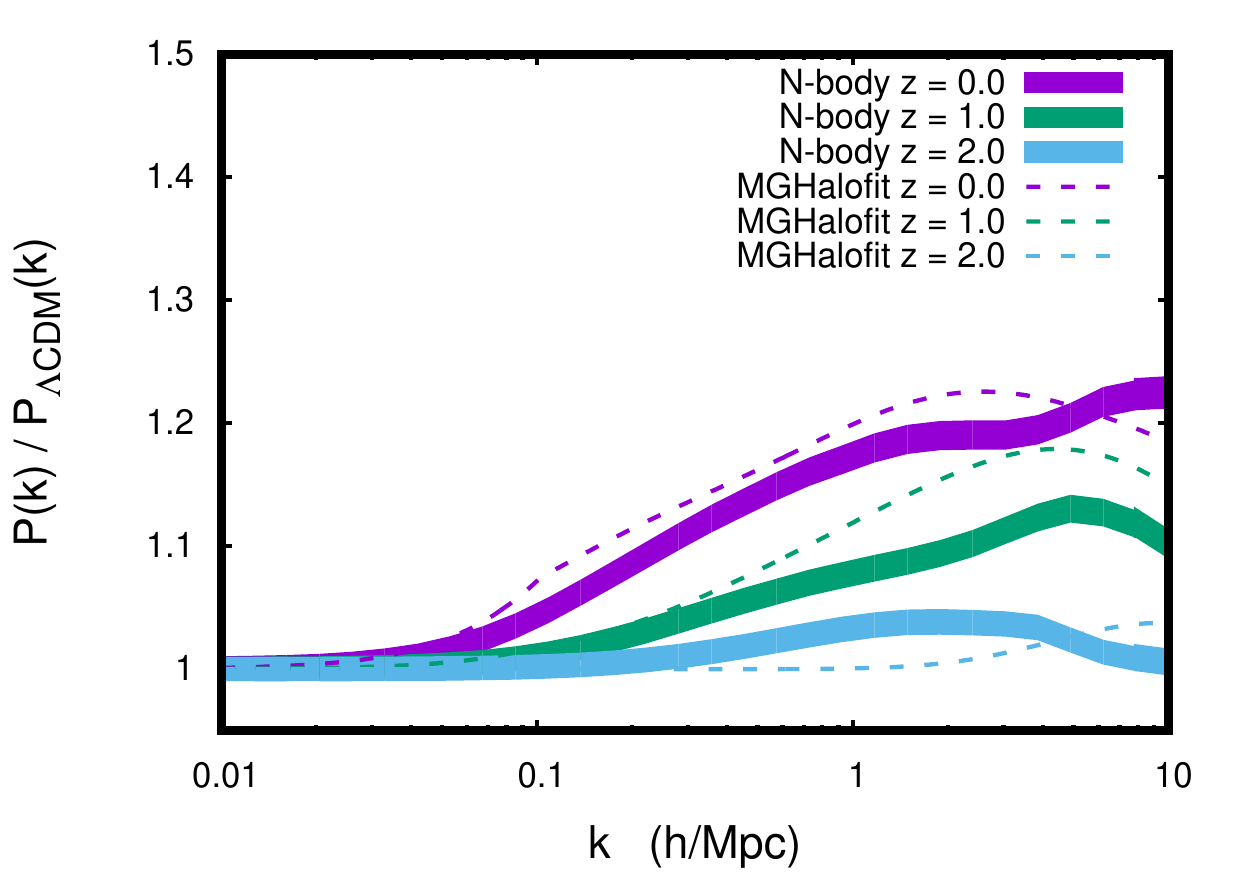}
\includegraphics[width=0.99\columnwidth]{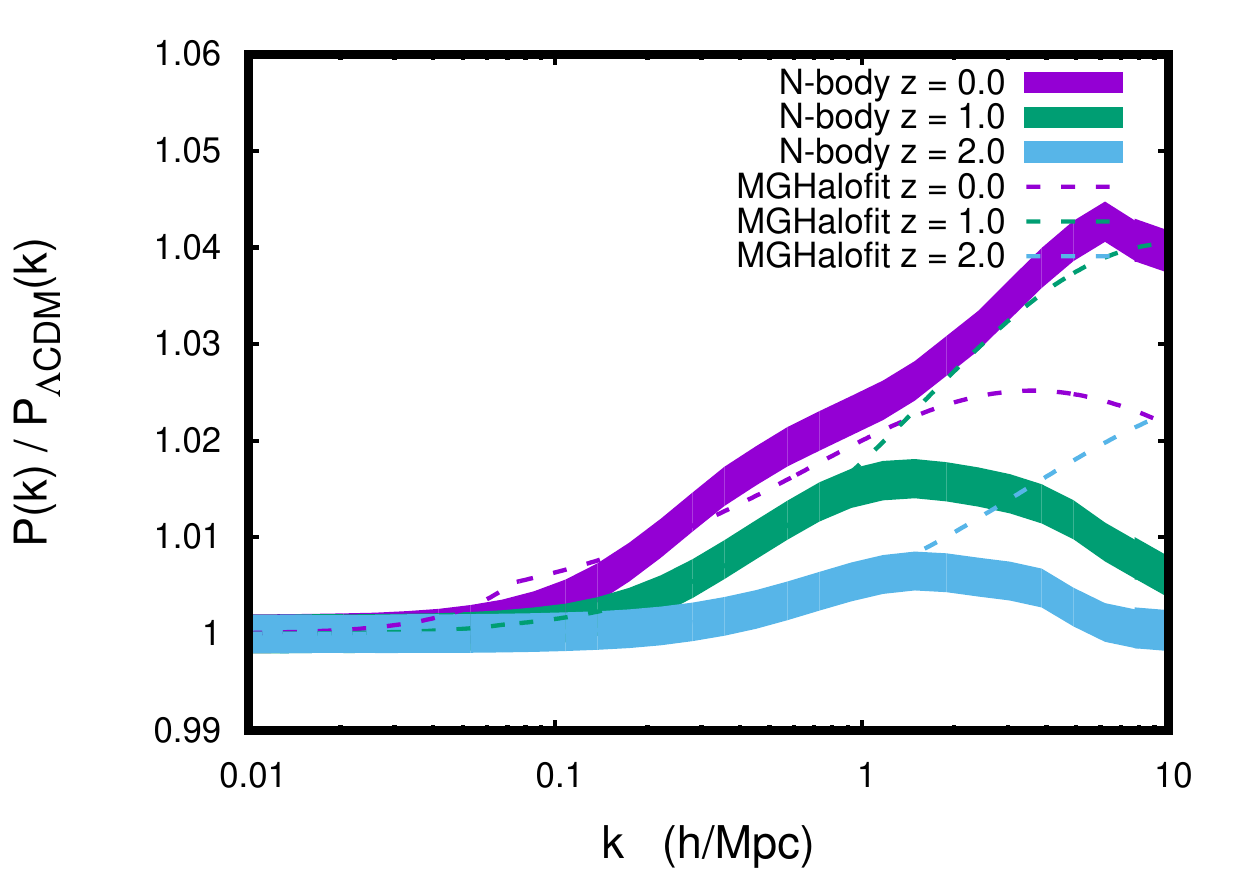}
\caption{Comparison of \mghalofit to our {\it N}-body data for $|f_{R0}| = 10^{-5}$ (left) and $10^{-6}$ (right). The shaded region correspond to $\pm$ 1\% (left) and $\pm 0.2\%$ (right) of the central value.}
\label{fig:mghalofit}
\end{figure*}

A fitting function, based on \halofit, for the Hu-Sawicki model already exists in the literature, namely \mghalofit. In \refFig{fig:mghalofit} we show a comparison of our fitting formula to \mghalofit which shows that our fit performs better.

\section{Forecast}\label{sec:forecasts}

In this section we will give an example application of the fitting formula presented in this paper by using it to compute forecasts for how well the Hu-Sawicki model can be constrained in future surveys (see \cite{2017PhRvD..95f3502A} for forecasts for general scalar-tensor theories). We adopt the Fisher-matrix formalism to the main cosmological observables for next-generation galaxy surveys, namely Galaxy Clustering (GC) and Weak Lensing (WL). The matter power spectrum is computed using \camb with the \halofit prescription to get the non-linear power spectrum for $\lcdm$ and then use our fitting-formula to go from $\lcdm$ to $f(R)$. We neglect cross-correlations between GC and WL. To perform our forecasts we use the survey parameters for a Euclid-like mission \cite{2011arXiv1110.3193L}.

For GC the main observable is the observed (redshift-space) galaxy power spectrum which we model as
\begin{align}
P_g(k,\mu,z) = \frac{D^2_{A,f}(z)/H_f(z)}{D^2_{A}(z)/H(z)}(b(z) + f(z)\mu^2)^2\times\nonumber\\
\times e^{-k^2\mu^2\sigma^2}P_{\rm DM}(k,z)
\end{align}
where $D_A$ is the angular diameter distance, $H$ is the Hubble function, $\mu$ is the cosine of the angle between the line of sight and the Fourier vector $\vec{k}$, $f(z)$ is the growth function and $\sigma^2 \equiv \sigma_r^2+\sigma_v^2$ is a parameter parametrising errors induced by spectroscopic redshift measurements ($\sigma_r$) and the fingers-of-god effect ($\sigma_v$) that we marginalise over. A subscript $_f$  denotes the value in the fiducial cosmology which we take to be a $\lcdm$ cosmology with parameters $n_s = 0.96$, $10^9A_s = 2.126$, $h = 0.67$, $\Omega_m = 0.32$, $\Omega_b = 0.05$, $\Omega_\Lambda = 0.68$, $m_\nu = 0.06$ eV and $\sigma_v = 300\text{ km/s}$. The fiducial $f(R)$ parameter is taken to be $|f_{R0}| = 10^{-5}$.

The Fisher matrix for GC is taken to be
\begin{align}
F_{ij} = \frac{V_{\rm survey}}{8\pi^2}\int_{-1}^1{\rm d}\mu \int_{k_{\rm min}}^{k_{\rm max}} {\rm d}k\left[\frac{\partial D}{\partial\theta_i}D^{-1}\frac{\partial D}{\partial\theta_j}D^{-1}\right]
\end{align}
where the data vector $D = P_g(k,\mu,z) + 1/n(z)$ with $V_{\rm survey}$ being the survey volume and $n(z)$ is the galaxy number density. We compute the constraints for two different values of $k_{\rm max} = 0.15$ and $0.25$ both with $k_{\rm min} = 0.008\,h/$Mpc. \newText{For computing the Fisher matrix we used $9$ $z$-bins in the range $z=0.95$ and $z=1.75$.}

The second probe we include is WL cosmic shear: the distortions in the ellipticities of galaxies due to bending of light around large cosmic structures. The cosmic shear is computed using $10$ redshift bins in redshift range $0 < z < 2.5$. The cosmic shear at a redshift bin $i$ is correlated with the cosmic shear at another redshift bin $j$ since light coming from each bin will propagate through some of the same structures on the way to us. The cross power spectrum of cosmic shear in bin $i$ and $j$ is determined by the underlying dark matter power spectrum via
\begin{align}
C_{ij}(\ell) = \frac{9}{4}\int_0^\infty {\rm d}z \frac{W_i(z)W_j(z)H^3(z)\Omega_m^2(z)}{(1+z)^4}\times\nonumber\\ \times P_{\rm DM}(k=\ell/r(z),z)
\end{align}
where $r(z)$ is the co-moving distance and $W$ is a a window function given by the photometric redshift distribution function and the galaxy number density distribution.

The Fisher matrix for WL is given by
\begin{align}
F_{ij} = f_{\rm sky}\sum_{\ell}^{\ell_{\rm max}} \frac{(2\ell + 1)\Delta \ell}{2}\text{tr}\left[\frac{\partial C}{\partial\theta_i}\text{Cov}^{-1}\frac{\partial C}{\partial\theta_j}\text{Cov}^{-1}\right]
\end{align}
where $C$ is a matrix with elements $C_{ij}$, $f_{\rm sky} = 0.36$ ($15000\text{ deg}^2$) is the sky-fraction covered by the survey, $\Delta \ell$ is the size of the $\ell$-bins, $\ell_{\rm max}$ is the maximum multipole number and $\text{Cov}$ is the WL covariance matrix. In this paper we consider the two values $\ell_{\rm max} = 1000$ and $3000$. \newText{We used $100$ logarithmically spaced $\ell$-bins between $\ell_{\rm min} = 100$ and $\ell_{\rm max}$ and we used $10$ evenly spaced $z$-bins between $z=0.15$ and $z=2.5$.} Apart from the particular numbers quoted above we use the same setup as \cite{Casas:2017eob} so see this paper for more details.

\begin{figure*}
\includegraphics[width=1.5\columnwidth]{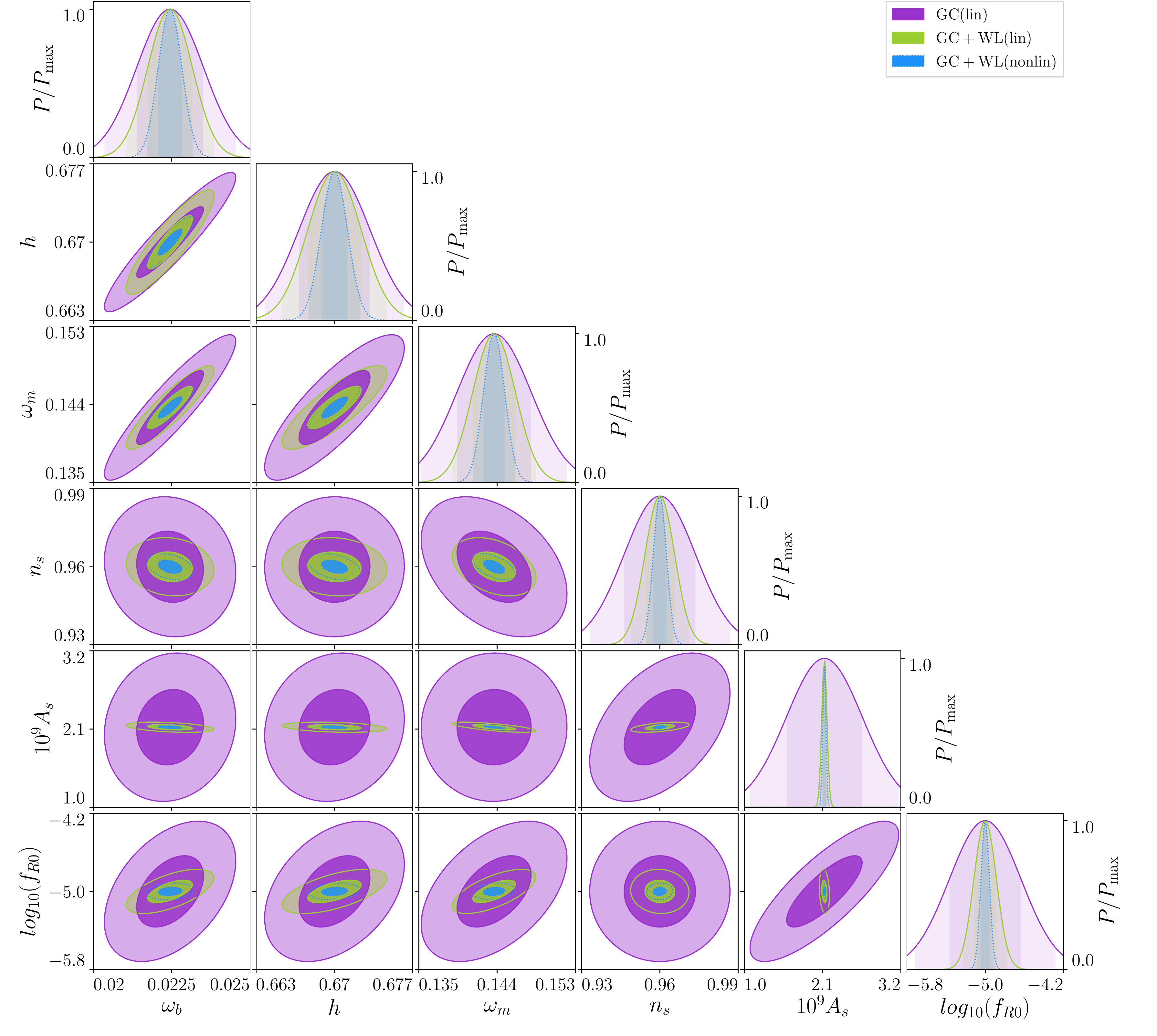}
\includegraphics[width=1.5\columnwidth]{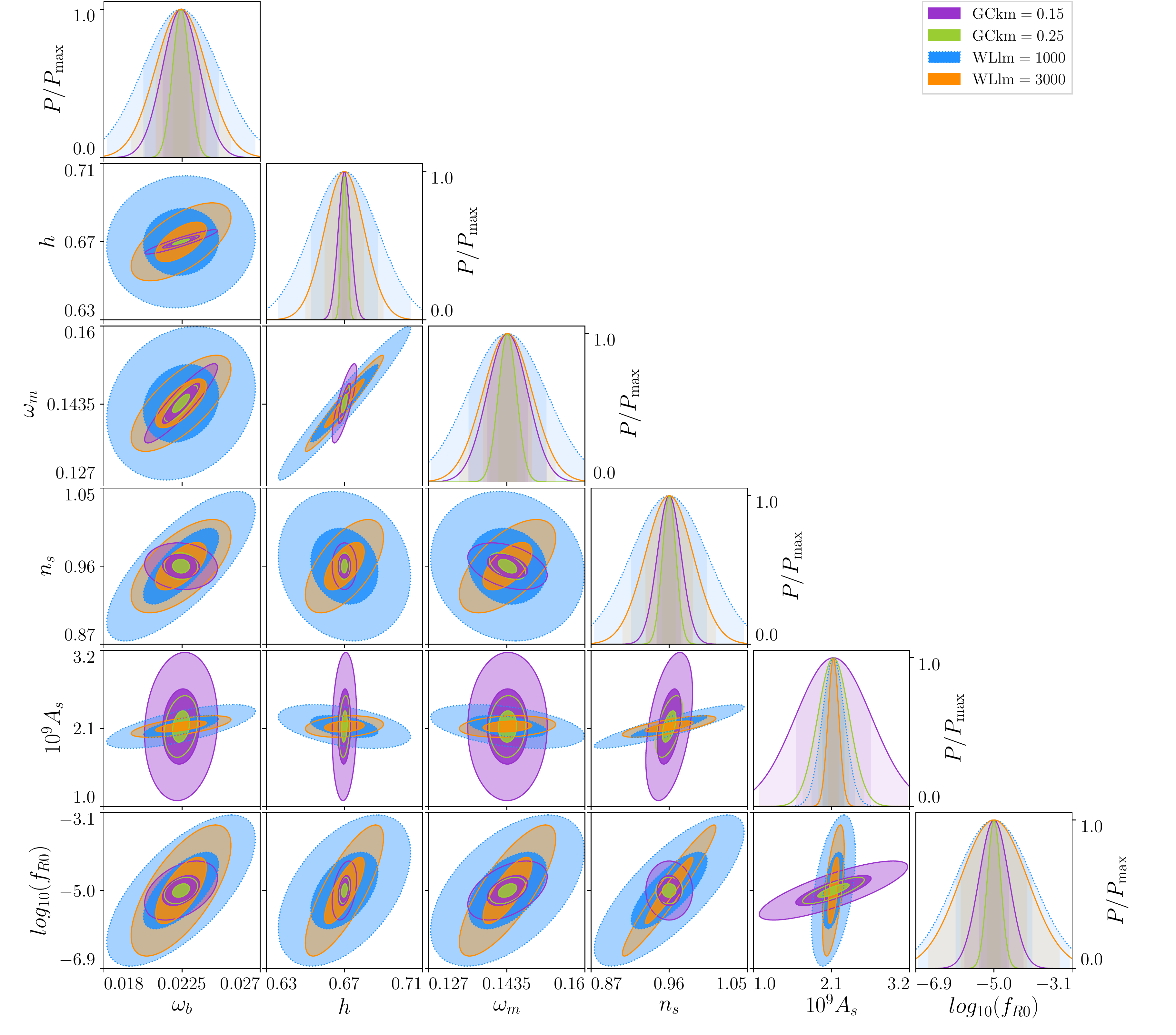}
\caption{Forecasted constraints on the Hu-Sawicki model, for the fiducial value $|f_{R0}| = 10^{-5}$, from galaxy clustering (GC) and weak-lensing (WL) in a Euclid-like survey for different values of $k_{\rm max}$ and $\ell_{\rm max}$. In the figure above (lin) refers to $k_{\rm max} = 0.15\hMpc$, $\ell_{\rm max } =1000$ and (nonlin) refers to $k_{\rm max} = 0.25\hMpc$, $\ell_{\rm max } =3000$.}
\label{fig:constraints}
\end{figure*}

In \refFig{fig:constraints} we show the Fisher forecast constraints we obtained using the fiducial value $|f_{R0}| = 10^{-5}$, which is slightly below the best constraints from current cosmological data (see e.g. \cite{2014JCAP...03..046D}). We show how the results change when going from only using fairly linear scales to including more and more non-linear scales ($k_{\rm max} = 0.15\hMpc,\ell_{\rm max} = 1000$ versus $k_{\rm max} = 0.25\hMpc,\ell_{\rm max} = 3000$) in the forecast. In \refTable{table:constraints} we show the marginalized constraints on the cosmological parameters for the different cases we have considered.

GC and WL are individually able to constrain $\log_{10}|f_{R0}|$ to $\sim 5\%$ and $\sim 15\%$ respectively depending on how many non-linear modes we include in the forecast. The modified gravity parameter $f_{R0}$ is seen to be mostly degenerate with the clustering amplitude $A_s$. Combining GC and WL we are able to break this degeneracy and bring the potential constraints down to $\sim 1-2\%$ ($\Delta f_{R0} \lesssim 2\cdot 10^{-6}$). However we caution that this is a simplified forecast not taking baryonic effects on the matter power spectrum or including massive neutrinos which both are known to be degenerate with a potential modified gravity signal \cite{2014MNRAS.440...75B,2019MNRAS.483..790A}.

\begin{table*}
\begin{tabular}{cccccc}
\hline
\multicolumn{6}{c}{GC $k_{\rm max} = 0.15\hMpc$}\\
\hline
$\sigma_{\Omega_b} = 4.55\%$ & $\sigma_h = 0.45\%$ & $\sigma_{\Omega_m} = 2.78\%$ & $\sigma_{n_s} = 1.04\%$ & $\sigma_{10^9A_s} = 23.81\%$ & $\sigma_{\log|f_{R0}|} = 8.00\%$ \\
\hline
\multicolumn{6}{c}{GC $k_{\rm max} = 0.25\hMpc$}\\
\hline
$\sigma_{\Omega_b} = 0.23\%$ & $\sigma_h = 0.15\%$ & $\sigma_{\Omega_m} = 1.39\%$ & $\sigma_{n_s} = 0.83\%$ & $\sigma_{10^9A_s} = 9.52\%$ & $\sigma_{\log|f_{R0}|} = 4.00\%$ \\
\hline
\multicolumn{6}{c}{WL $\ell_{\rm max} = 1000$}\\
\hline
$\sigma_{\Omega_b} = 9.09\%$ & $\sigma_h = 2.99\%$ & $\sigma_{\Omega_m} = 5.56\%$ & $\sigma_{n_s} = 4.17\%$ & $\sigma_{10^9A_s} = 9.52\%$ & $\sigma_{\log|f_{R0}|} = 18.00\%$ \\
\hline
\multicolumn{6}{c}{WL $\ell_{\rm max} = 3000$}\\
\hline
$\sigma_{\Omega_b} = 4.55\%$ & $\sigma_h = 1.49\%$ & $\sigma_{\Omega_m} = 3.47\%$ & $\sigma_{n_s} = 3.12\%$ & $\sigma_{10^9A_s} = 3.76\%$ & $\sigma_{\log|f_{R0}|} = 16.00\%$ \\
\hline
\multicolumn{6}{c}{GC+WL $k_{\rm max} = 0.15 \hMpc$, $\ell_{\rm max} = 1000$}\\
\hline
$\sigma_{\Omega_b} = 3.12\% $ & $\sigma_h = 0.3\%$ & $\sigma_{\Omega_m} = 1.39\%$ & $\sigma_{n_s} = 0.62\%$ & $\sigma_{10^9A_s} = 1.88\%$ & $\sigma_{\log|f_{R0}|} = 2.00\%$ \\
\hline
\multicolumn{6}{c}{GC+WL $k_{\rm max} = 0.25 \hMpc$, $\ell_{\rm max} = 3000$}\\
\hline
$\sigma_{\Omega_b} = 1.79\%$ & $\sigma_h = 0.15\%$ & $\sigma_{\Omega_m} = 0.69\%$ & $\sigma_{n_s} = 0.31\%$ & $\sigma_{10^9A_s} = 0.94\%$ & $\sigma_{\log|f_{R0}|} = 0.80\%$ \\
\hline
\end{tabular}
\caption{Constraints on the parameters in the Hu-Sawicki model, for the fiducial value $|f_{R0}| = 10^{-5}$, coming from galaxy clustering (above), weak-lensing (middle) and combined (below) for two different values of $k_{\rm max}$ and $\ell_{\rm max}$.}
\label{table:constraints}
\end{table*}

\section{Conclusions}\label{sec:conc}

In this paper we have presented a fitting function for the linear and non-linear matter power spectrum of the Hu-Sawicki $f(R)$ model using power spectra computed from {\it N}-body simulations for several different values of the model parameters. This is one of the most studied modified gravity models and often the fiducial choice when trying to constrain modified gravity effects in observational data.

We have shown that the enhancement has a weak cosmology dependence which allows us to make the fit for a fixed cosmology. Any cosmology dependence can be accurately included by using inexpensive tools like linear theory, the halo model or COLA simulations.

We have demonstrated that the fitting function is accurate over a large range of scales and redshifts. We also tested it against the existing \mghalofit code and found that our fitting function generally performs better.

One can easily integrate our fitting function in any approach that produces a non-linear matter power spectrum for $\lcdm$. With this paper we provide the fitting function already implemented\footnote{This can be found at https://github.com/HAWinther/FofrFittingFunction} in many common programming languages like C, Fortran and Python plus an implementation in both \camb and \class.

Finally, as an application, we have used the fitting functions to compute Fisher forecasts for how well a Euclid-like survey will be at constraining the Hu-Sawicki model. We find that the potential constraints from combining GC and WL from a Euclid-like survey, when including a reasonable amount of non-linear scales in the forecast, are at the $\Delta f_{R0} \sim 2\cdot 10^{-6}$ level. This is without taking into account baryonic effects in the power spectrum and including massive neutrinos, but nevertheless shows the potential constraining power in future survey when including non-linear scales.

The same approach as used in this paper can likely be applied to cheaply create emulators for other non-standard gravity models, possibly in conjunction with the semi-analytical method of \cite{2018arXiv181205594C}.

\section*{Acknowledgment}
KK and HAW are supported by the European Research Council through 646702 (CosTesGrav). KK is also supported by the UK Science and Technologies Facilities Council grants ST/N000668/1. GBZ is supported by NSFC Grant No.  11673025, and by a Royal Society-Newton Advanced Fellowship. LL is supported by a Swiss National Science Foundation Professorship grant (No.~170547). SC acknowledges support from CNRS and CNES grants. BL acknowledge the support of the UK STFC Consolidated Grants (ST/P000541/1 and ST/L00075X/1) and Durham University. BL is also supported in part by the European Union’s Horizon 2020 research and innovation program. MB acknowledges support from the Italian Ministry for Education, University and Research (MIUR) through the SIR individual grant SIMCODE (project number RBSI14P4IH), from the grant MIUR PRIN 2015 "Cosmology and Fundamental Physics: illuminating  the  Dark Universe  with Euclid", and from the agreement ASI n.I/023/12/0 “Attività relativealla fase B2/C per la missione Euclid”.

\newpage
$\left.\right.$
\newpage
$\left.\right.$

\bibliography{bibfile}

\end{document}